\documentclass[12pt]{article}

\usepackage{amsmath, amssymb, amsthm, mathtools, bm} 
\usepackage{graphicx, subcaption, float}             
\usepackage{booktabs, multirow, longtable, array, threeparttable, makecell, placeins}    
\usepackage{rotating}
\usepackage{tablefootnote}
\usepackage[
  backend=biber,
  style=authoryear-comp,  
  maxcitenames=1,    
  mincitenames=1,    
  maxbibnames=99,    
  uniquelist=false,  
  uniquename=false,   
  doi=false,
  url=false,
  eprint=false,
  isbn=false,
]{biblatex}

\DeclareFieldFormat[article]{volume}{\mkbibbold{#1}}
\renewbibmacro*{volume+number+eid}{%
  \setunit{\addcomma\space}%
  \printfield[volume]{volume}%
  \setunit{}%
  \printfield[parens]{number}%
  \setunit{\addcomma\space}%
  \printfield{eid}}
\usepackage{url, xcolor, setspace, geometry, float, titlesec, authblk}         

\usepackage[colorlinks=true,
            linkcolor=blue,
            citecolor=blue,
            urlcolor=blue]{hyperref} 

\titleformat{\subsection}[block] 
  {\normalfont\fontsize{12}{12}\bfseries\itshape} 
  {} 
  {0pt} 
  {}

\title{Refining capture-recapture methods to estimate case counts in a finite population setting}

\author[1]{Michael Doerfler}
\author[1]{Wenhao Mao}
\author[2]{Lin Ge}
\author[3]{Yuzi Zhang}
\author[4]{Timothy L. Lash}
\author[4]{Kevin C. Ward}
\author[1]{Lance A. Waller}
\author[1]{Robert H. Lyles}

\affil[1]{Department of Biostatistics and Bioinformatics, Rollins School of Public Health, Emory University, Atlanta, Georgia, USA}
\affil[2]{Department of Epidemiology and Biostatistics, School of Public Health, Indiana University, Bloomington, Indiana, U.S.A.}

\affil[3]{Division of Biostatistics, College of Public Health, Ohio State University, Columbus, Ohio, U.S.A}

\affil[4]{Department of Epidemiology, Rollins School of Public Health, Emory University, Atlanta, Georgia, U.S.A.}

\date{}

\begin{document}

\maketitle

\begin{abstract}
     In this paper, we expand upon and refine a monitoring strategy proposed for surveillance of diseases in finite, closed populations. This monitoring strategy consists of augmenting an arbitrarily non-representative data stream (such as a voluntary flu testing program) with a random sample (referred to as an “anchor stream”). This design allows for the use of traditional capture-recapture (CRC) estimators, as well as recently proposed anchor stream estimators that more efficiently utilize the data.  Here, we focus on a particularly common situation in which the first data stream only records positive test results, while the anchor stream documents both positives and negatives.  Due to the non-representative nature of the first data stream, along with the fact that inference is being performed on a finite, closed population, there are standard and non-standard finite population effects at play.  Here, we propose two methods of incorporating finite population corrections (FPCs) for inference, along with an FPC-adjusted Bayesian credible interval.  We compare these approaches with existing methods through simulation and demonstrate that the FPC adjustments can lead to considerable gains in precision.  Finally, we provide a real data example by applying these methods to estimating the breast cancer recurrence count among Metro Atlanta-area patients in the Georgia Cancer Registry-based Cancer Recurrence Information and Surveillance Program (CRISP) database.
\end{abstract}


\section{Introduction}

Capture-recapture (CRC) methods have a long history and have been utilized in a wide range of fields. These approaches were originally developed for ecological problems involving the estimation of the size of wildlife populations based on repeated capture and release efforts (\cite{Petersen1896}, \cite{Lincoln1930}).  Recently, CRC methods have also seen an increase in popularity in other settings, such as for monitoring infectious diseases, assessing the magnitude of modern day slavery, and estimating the prevalence of illicit drug use (\cite{Boehning2017}, \cite{Abeni1994}, \cite{Silverman2020}, \cite{Gemmell2004}).  Fundamentally, CRC methods involve analyzing multiple overlapping data streams in order to conduct inference on a population.  This population can be considered open or closed depending on whether we allow for movement into and out of the population. In this paper, we focus on closed populations, in which estimation of the total population size is typically the primary question of interest (for examples of open population CRC methodologies, see \cite{Jolly1965}, \cite{Pollock1982}, \cite{Pradel1996}, \cite{Schwarz1996}).  In these closed population settings, for T data streams, the capture experience of each individual in the population can be represented by a vector of 1’s and 0’s of length T, with a 1 indicating the individual had been captured in the corresponding capture effort and a 0 meaning it had not.   Importantly, the number of individuals who have never been captured (i.e., a vector of all 0’s) is unknown.  Estimating this unknown count is the critical problem in inferring the total population size N.  

Over the years, numerous CRC methods have been proposed for dealing with this problem. A special class of estimators, known as the Lincoln-Petersen and Chapman estimators, can be used in the two-stream setting when the so-called LP conditions are met (\cite{Seber1982}).  These conditions require that the two streams be independent at the population-level (see \cite{Chao2008} and \cite{Lyles2021} for distinctions between individual-level and population-level models).  If this requirement is met, then these classical estimators can be used.  Unfortunately, the independence assumption is often difficult to meet, and numerous estimators have been proposed for situations in which the LP conditions are invalid.

However, a study design in which the LP conditions are valid is when one data stream is implemented agnostically of the other as a simple random sample.  This has been referred to as the “anchor stream design” (with the random sample being called the “anchor stream”) (\cite{Lyles2022, Lyles2023}; \cite{Ge2023}).  This design not only validates the Lincoln-Petersen and Chapman estimators but also unlocks novel estimators that allow for more efficient use of the entire data.  These estimators are developed without any assumptions about or restrictions on the sampling mechanism behind the non-anchor stream; that is, the non-anchor stream can be arbitrarily non-representative.

In this study, we expand upon the use of the anchor stream design for monitoring disease in a fixed, finite population of known total size $N_{tot}$, such as a retirement community or a registry cohort.  In particular, we seek to refine inferences under what we consider to be the most common anchor stream scenario.  In this scenario, the non-anchor stream only records positive test results (hereafter, the non-anchor stream is referred to as Stream 1. We will also occasionally refer to the anchor stream as Stream 2).  That is, those who are sampled in Stream 1 and test negative and those who are not sampled in Stream 1 at all are indistinguishable in the data.  We refer to this scenario as the “5-cell case" since all members of the population can be represented as falling into one of five categories (Table 3).  This case should be distinguished from the “7-cell case” (Table 1), in which both Stream 1 and Stream 2 record negative results, and the “4-cell case” (Table 2) in which neither Stream 1 nor Stream 2 records negative results.


\begin{table}[H]
\centering
\caption{The 7-cell case in which both Stream 1 and Stream 2 record negative and positive test results.}
\label{tab:7cell}
\begin{tabular}{ll}
\toprule
\textbf{Cell Count} & \textbf{Observation Type} \\
\midrule
$n_{1}$ & Sampled in both streams, tested negative. \\
$n_{2}$ & Sampled in both streams, tested positive. \\
$n_{3}$ & Sampled in Stream 1 but not 2, tested negative. \\
$n_{4}$ & Sampled in Stream 1 but not 2, tested positive. \\
$n_{5}$ & Sampled in Stream 2 but not 1, tested negative. \\
$n_{6}$ & Sampled in Stream 2 but not 1, tested positive. \\
$n_{7}$ & Not sampled in either stream. \\
\bottomrule
\end{tabular}
\end{table}


\begin{table}[H]
\centering
\caption{The 4-cell case in which neither Stream 1 nor Stream 2 records negative test results.}
\label{tab:3cell}
\begin{tabular}{l p{10cm}}
\toprule
\textbf{Cell Count} & \textbf{Observation Type} \\
\midrule
$n_{2}$ & Sampled in both streams, tested positive. \\
$n_{4}$ & Sampled in Stream 1 but not 2, tested positive. \\
$n_{6}$ & Sampled in Stream 2 but not 1, tested positive. \\
$n_{1357}$ & Not sampled in either stream \textbf{OR} Sampled in Stream 1 but not 2, tested negative \textbf{OR} Sampled in Stream 2 but not 1, tested negative \textbf{OR} Sampled in both streams, tested negative.\\
\bottomrule
\end{tabular}
\end{table}


\begin{table}[H]
\centering
\caption{The 5-cell case in which Stream 2 records both positive and negative test results, while Stream 1 only records positive test results.}
\label{tab:5cell}
\begin{tabular}{ll}
\toprule
\textbf{Cell Count} & \textbf{Observation Type} \\
\midrule
$n_{15}$ & Sampled in both streams, tested negative \textbf{OR} \\
         & Sampled in Stream 2 but not 1, tested negative. \\
$n_{2}$  & Sampled in both streams, tested positive. \\
$n_{4}$  & Sampled in Stream 1 but not 2, tested positive. \\
$n_{6}$  & Sampled in Stream 2 but not 1, tested positive. \\
$n_{37}$ & Not sampled in either stream \textbf{OR} \\
         & Sampled in Stream 1 but not 2, tested negative. \\
\bottomrule
\end{tabular}
\end{table}

Note that when Stream 1 does not record negative results, cells 1 and 5, and cells 3 and 7 are indistinguishable in Table 1 and thus are consolidated in Table 3. We assume that the case status of all individuals is correctly recorded with no false positives or false negatives (see \cite{Ge2023, Ge2024} for anchor stream analysis with corrections for misclassification).  Interest is in inference on $N$, the number of disease cases in the cohort.

To make these ideas more concrete, we consider both a hypothetical scenario and a real-data example.

\subsection{Hypothetical Scenario: Serosurveys for Cumulative Flu Incidence}

Monitoring of cumulative influenza cases in a particular flu season is critical in assessing disease burden and evaluating vaccine efficacy.  Serosurveys allow for the detection of past infection by detecting particular antibodies in a patient’s blood (\cite{Haselbeck2022}).  Importantly, serological tests indicate infection history and are independent of current infection status (\cite{Haselbeck2022}).  In this scenario, the goal is to estimate the cumulative number of flu cases in a particular geographic region over the course of a given flu season.  Stream 1 is a voluntary testing system available throughout the season that members of the population can enroll in to test for the presence of infection.  At the end of the flu season, a random sample of the population is taken and administered a serosurvey, such as a hemagglutination inhibition (HAI) assay (\cite{Kittikraisak2024}).  This serosurvey documents both positive and negative results and constitutes the anchor stream.  Since those who test negative in Stream 1 could theoretically still become infected by the end of the season, negative test results in Stream 1 cannot be used for determining whether an infection took place over the study period, meaning this would be an example of the 5-cell situation.

\subsection{Real-Data Example: CRISP Cancer Recurrence}

The Cancer Recurrence and Surveillance Program (CRISP) was established in conjunction with the Georgia Cancer Registry (GCR) to monitor and register recurrences among Georgia cancer survivors.  In our study, similarly to \cite{Ge2023}, we focus on GCR breast cancer survivors for whom initial diagnosis and treatment occurred in one of five metro Atlanta hospitals and whose residential addresses have remained in the Atlanta area.  Stream 1 consists of validated Georgia Commission on Cancer (CoC) recurrence reports. These reports have been identifying potential recurrences since 2013 (\cite{Ge2023}).  The anchor stream consists of a random sample of 200 breast cancer patients that was drawn in 2022, with CRISP abstractors reviewing medical records to determine whether a true recurrence had occurred, following the same validation protocol as for the CoC reports. With the assumption that the CoC validated records are accurate, these data can be analyzed under the 5-cell scenario to validly estimate the number of breast cancer recurrences among Atlanta patients in the GCR. \\

Point estimation for $N$ under the 5-cell case was briefly introduced in \cite{Lyles2022}. However, details on inference, and in particular, how to adjust for finite population effects, were not discussed. When estimating disease prevalence in finite population settings, inferences made without appropriate adjustments can be extremely imprecise (\cite{Cochran}).  In this paper, we propose an FPC-adjusted variance estimator to accompany point estimation of $N$, as well as an FPC-adjusted Bayesian credible interval approach.  We compare these approaches to non-FPC-adjusted inferences and show that our adjustments can lead to considerable gains in precision.

The rest of this paper is organized as follows.  In Section 2 we introduce FPC-adjusted variance estimates for the 5-cell setting, as well as the FPC-adjusted adapted Bayesian credible interval approach.  Section 3 contains the results of simulation studies implemented to compare FPC-adjusted to non-FPC-adjusted inferences, as well as different competing CRC estimators.  Additionally, Section 3 contains the results of a real-data analysis in which we apply our methods to estimate the number of breast cancer recurrences in the CRISP cohort.  Finally, in Section 4 we discuss these results and provide concluding remarks.

\section{Methods}

The total size of the fixed, finite population is given by $N_{\text{tot}}$, and interest is in inference on the number of individuals in this population who are diseased, denoted by $N$. Under the anchor stream design, a number of estimators are validated. The most basic estimator uses only the information from the anchor stream random sample of size $n_{\text{RS}}$. This is

\[
\widehat{N}_{\text{RS}} = N_{\text{tot}} \widehat{\pi}_{\text{RS}},
\]

\noindent where $
\widehat{\pi}_{\text{RS}} = \frac{n_{\text{RS}}^+}{n_{\text{RS}}} = \frac{n_2 + n_6}{n_{\text{RS}}}, \quad 
\widehat{\text{Var}}\left(\widehat{N}_{\text{RS}}\right) = N_{\text{tot}}^2 \widehat{\text{Var}}\left(\widehat{\pi}_{\text{RS}}\right),
$ and $n_{\text{RS}}^+$ is the number of subjects that tested positive for disease in the random sample. The FPC-adjusted variance of $\widehat{\pi}_{\text{RS}}$ is provided by \cite{Cochran}, with

\[
\widehat{\text{Var}}\left(\widehat{\pi}_{\text{RS}}\right) = \text{FPC} \cdot \frac{\widehat{\pi}_{\text{RS}} (1 - \widehat{\pi}_{\text{RS}})}{n_{\text{RS}}},
\]

\noindent where $
\text{FPC} = \frac{n_{\text{RS}}(N_{\text{tot}} - n_{\text{RS}})}{N_{\text{tot}}(n_{\text{RS}} - 1)}.
$

An FPC-adjusted Wald confidence interval is commonly reported alongside $\widehat{N}_{\text{RS}}$. Recently, an FPC-adjusted Bayesian credible interval approach has been introduced as an alternative (\cite{Lyles2023}).  

Since the LP conditions automatically hold under the anchor stream design, the Lincoln-Petersen and Chapman estimators are both validated as well. The Chapman estimator is a bias-reduced version of the Lincoln-Petersen estimator, with

\[
\widehat{N}_{\text{Chap}} = \frac{(n_2 + n_4 + 1)(n_2 + n_6 + 1)}{n_2 + 1} - 1,
\]

\noindent and $
\widehat{\text{Var}}\left(\widehat{N}_{\text{Chap}}\right) = \frac{(n_2 + n_4 + 1)(n_2 + n_6 + 1)(n_4)(n_6)}{(n_2 + 1)^2 (n_2 + 2)}
$ (\cite{Seber1982}; \cite{Chapman1951}; \cite{Wittes1972}). As in \cite{Lyles2021}, we prefer to report a logit-based confidence interval to accompany this estimator (\cite{Sadinle2009}).  

While these estimators are all valid under the anchor stream design, none are fully efficient. In the most basic anchor stream scenario (the 4-cell case), one can show that under a population-level multinomial model, the maximum likelihood estimator (MLE) of $N$ is given by

\begin{equation}
\widehat{N}_{4,\psi} = n_2 + n_4 + \frac{n_6}{\psi}, 
\end{equation} where $\psi$ is the known sampling probability into the anchor stream (\cite{Chen2020}). This estimator was originally utilized for sensitivity and uncertainty analysis for capture-recapture (\cite{Zhang2023SensitivityUncertaintyAnalysis}). A corresponding variance estimate (based on the multivariate delta-method) is

\begin{equation}
\widehat{\text{Var}}\left(\widehat{N}_{4,\psi}\right) = \frac{n_6 (1 - \psi)}{\psi^2}.
\end{equation} A standard Wald interval can be reported alongside (1). An adjusted Bayesian credible interval approach has also been introduced (\cite{Lyles2023}).  

In the 7-cell scenario, the MLE of $N$ under a population-level multinomial model is given by $
\widehat{N}_7 = n_2 + n_4 + n_6 \left(\frac{n_5 + n_6 + n_7}{n_5 + n_6}\right)
$.  \cite{Lyles2023} discuss inference in the 7-cell case.

\subsection{5-Cell Case}

 In this paper, we focus on the five-cell case, which would apply to both the CRISP data and the hypothetical flu monitoring scenario.  We adopt a population-level multinomial model (Table 4), with parameters in Table 4 defined by $\psi=Pr\left(Sampled\ in\ Stream\ 2\right)$, $\phi=Pr\left(Sampled\ in\ Stream\ 1\right)$, $\pi_{S1}=Pr\left(Diseased\middle| S a m p l e d\ in\ Stream\ 1\right)$, and $\pi_{\overline{S}1}=Pr(Diseased\ |\ Not\ Sampled\ in\ Stream\ 1)$.  Not all parameters in this model are identifiable, so we reparametrize by defining parameters $\theta=\pi_{S1}\phi$ and $\pi=\pi_{S1}\phi+\pi_{\overline{S}1}\left(1-\phi\right).$


\begin{table}[ht]
\centering
\caption{5-cell case with population-level multinomial likelihood contributions.}
\label{tab:5cell_likelihood_wrapped}
\begin{tabular}{p{1.5cm} p{6cm} p{7cm}}
\toprule
\textbf{Cell Count} & \textbf{Observation Type} & \textbf{Likelihood Contribution} \\
\midrule

$n_{15}$ &
\parbox[t]{6cm}{Sampled in both streams, tested negative \textbf{OR} sampled in Stream 2 but not 1, tested negative.} &
\parbox[t]{7cm}{\raggedright
\(\begin{aligned}[t]
p_{15} &= \psi (1 - \pi_{S1}) \phi + \psi (1 - \pi_{\overline{S}1})(1 - \phi) \\
       &= \psi (1 - \pi)
\end{aligned}\)
} \\ \hline

$n_2$ &
\parbox[t]{6cm}{Sampled in both streams, tested positive.} &
\parbox[t]{7cm}{\raggedright
\(\begin{aligned}[t]
p_2 &= \psi \pi_{S1} \phi \\
    &= \psi \theta
\end{aligned}\)
} \\ \hline

$n_4$ &
\parbox[t]{6cm}{Sampled in Stream 1 but not 2, tested positive.} &
\parbox[t]{7cm}{\raggedright
\(\begin{aligned}[t]
p_4 &= (1 - \psi)\,\pi_{S1}\,\phi \\
    &= (1 - \psi)\,\theta
\end{aligned}\)
} \\ \hline

$n_6$ &
\parbox[t]{6cm}{Sampled in Stream 2 but not 1, tested positive.} &
\parbox[t]{7cm}{\raggedright
\(\begin{aligned}[t]
p_6 &= \psi\,\pi_{\overline{S}1}\,(1 - \phi) \\
    &= \psi\,(\pi - \theta)
\end{aligned}\)
} \\ \hline

$n_{37}$ &
\parbox[t]{6cm}{Not sampled in either stream \textbf{OR} sampled in Stream 1 but not 2, tested negative.} &
\parbox[t]{7cm}{\raggedright
\(\begin{aligned}[t]
p_{37} &= (1 - \psi)(1 - \pi_{S1}) \phi + (1 - \psi)(1 - \phi) \\
       &= (1 - \psi)(1 - \theta)
\end{aligned}\)
} \\

\bottomrule
\end{tabular}
\end{table}

\noindent It can be shown that the MLE of $\pi$ (the prevalence) is given by,
\begin{equation} \widehat{\pi}_5=\frac{1}{N_{tot}}\left(n_2+n_4+n_6 \left(\frac{n_{15}+n_6+n_{37}}{n_{15}+n_6}\right)\right). \end{equation}

\noindent The MLE of N then follows naturally as,
\begin{equation} \widehat{N}_5=n_2+n_4+n_6\left(\frac{n_{15}+n_6+n_{37}}{n_{15}+n_6}\right). \end{equation}

We derive a corresponding variance estimator for (3) via the multivariate delta method, which is given by
\begin{equation} \widehat{Var}\left(\widehat{\pi}_5\right)_{unadj}=d^\prime\Sigma d,  \end{equation}
where $d={\left(\frac{-\widehat{p}_6\widehat{p}_{37}}{\left(\widehat{p}_{15}+\widehat{p}_6\right)^2},1,1,1+\frac{\widehat{p}_{37}\widehat{p}_{15}}{\left(\widehat{p}_{15}+\widehat{p}_6\right)^2},\frac{\widehat{p}_6}{\widehat{p}_{15}+\widehat{p}_6}\right)\ }^T$, $\widehat{p}_k=\frac{n_k}{N_{tot}}$, and $\Sigma$ is a 5 by 5 matrix with $\Sigma_{ij}=\frac{\widehat{p}_i\left(1-\widehat{p}_i\right)}{N_{tot}}$ if $i=j$ and $\Sigma_{ij}=\frac{-\left(\widehat{p}_i\right)\left(\widehat{p}_j\right)}{N_{tot}}$ if $i\ \neq j.$  An accompanying variance estimator for $\widehat{N}_5$ would then be,
\begin{equation}
\widehat{Var}(\widehat{N}_5)_{unadj}=N_{tot}^2\widehat{Var}\left(\widehat{\pi}_5\right)_{unadj}.
\end{equation}
We refer to (6) as the “unadjusted variance estimator.” To address division by 0 issues, in the rare case that one of the five cell counts is 0, we replace each $\widehat{p}_k$ in (5) and (6) by $\frac{n_k+0.5}{N_{tot}+5(0.5)}$, which is the posterior mean of ${\bf p}=(p_{15}, p_{2}, p_{4}, p_{6}, p_{37})^{T}$ if a Jeffreys prior were imposed on {\bf p}.

\subsection{FPC Adjustments}

The need for a FPC in standard sampling problems is well recognized, as the nature of sampling from a finite population introduces effects that result in standard variance estimators being overly conservative (\cite{Cochran}, \cite{Lohr2021}). In the anchor stream design, previous work on the 7-cell case has noted that FPC effects can apply depending on the inferential goal (\cite{Lyles2023}). These effects are both standard and non-standard, due to the combination of simple random sampling in the anchor stream with the arbitrarily non-representative Stream 1.

We find that FPC effects can apply in the 5-cell case as well and need to be adjusted for depending on the inferential goals of the analyst. If the total population $N_{tot}$ is viewed as being a representative sample of an even larger superpopulation that we are interested in, then FPC adjustments are not needed. However, if the analyst is instead only interested in the number of diseased individuals in the finite, fixed population $N_{tot}$, without any concern about a larger superpopulation (such as in our motivating examples), then FPC adjustments would be necessary, as estimator (6) would result in inferences that are overly conservative (see simulation results).

Due to this, we propose an FPC-adjusted variance estimator to accompany (3) and (4).  This is based on observing that the estimator in (3) can be rewritten as follows:

\begin{equation}
    \hat{\pi}_5=w+(1-w)\hat{p}^{*}
\end{equation} where $w=\frac{n_2+n_4}{N_{tot}}$ and $\hat{p}^{*}=\frac{n_6}{n_{15}+n_6}$. We treat the weight $w$ as a constant for the purposes of variance estimation. Then, observing that $\widehat{p}^{*}$ can effectively be treated as a sample proportion computed from a random sample of size $n_{rs}^{*}=n_{15}+n_6$ drawn from a finite population of size $N_{tot}^{*}=n_{15}+n_6+n_{37}$, we can apply a standard FPC adjustment to derive $\widehat{Var}\left(\widehat{p}^{*}\right)_{FPC}=\ \frac{\left(n_{RS}^\ast\right)\left(N_{tot}^\ast-n_{RS}^\ast\right)}{\left(N_{tot}^\ast\right)\left(n_{RS}^\ast-1\right)}\frac{\widehat{p}^{*}\left(1-\widehat{p}^{*}\right)}{\left(n_{RS}^\ast\right)}$ (further details on this derivation can be found in the Appendix). This yields the following FPC-adjusted variance estimator for $\widehat{\pi}_5:$

\begin{equation}  
\widehat{Var}\left(\widehat{\pi}_5\right)_{FPC1}=\left(1-w\right)^2\widehat{Var}\left(\widehat{p}^{*}\right)_{FPC}.  
\end{equation}

\noindent The FPC-adjusted variance estimate for $\widehat{N_5}$ follows naturally:

\begin{equation}
\widehat{Var}\left(\widehat{N_5}\right)_{FPC1}=N_{tot}^2\widehat{Var}\left(\widehat{\pi_5}\right)_{FPC1}.
\end{equation}

A second FPC-adjusted variance estimator in which the weights are not treated as constants (see Appendix for derivation) is given by,

\begin{equation}
\widehat{Var}\left(\widehat{N}_5\right)_{FPC2}=N_{tot}^2\left(\frac{\left(1-\frac{n_6}{n_{15}+n_6}\right)^2(\frac{n_2+n_4}{N_{tot}})\left(1-\frac{n_2+n_4}{N_{tot}}\right)}{N_{tot}}\right)\ +\ \widehat{Var}\left(\widehat{N}_5\right)_{FPC1}. 
\end{equation} However, we find that inference based on (10) is generally conservative (though more precise than using the unadjusted variance estimate) and generally recommend using (9) (see simulation results).

\subsection{Adapted Bayesian Credible Interval}

We use an adapted Bayesian credible interval to accompany the point estimators (3) and (4), based on first applying a Dirichlet prior to the five capture probabilities ${\bf p}=\left(p_{15},\ldots,p_{37}\right)^T$  in the multinomial model.  In this approach, we first take $M$ posterior draws of ${\bf p}$ from a $Dirichlet\left(n_{15}+\frac{1}{2},n_2+\frac{1}{2},n_4+\frac{1}{2},n_6+\frac{1}{2},n_{37}+\frac{1}{2}\right)$ distribution.  Then, for each posterior draw ${\bf p_m}^{*},m=1,\ldots,M$, we calculate $\widehat{N}_{5,m}^{*}=N_{tot}\left(p_{2,m}^{*}+p_{4,m}^{*}+p_{6,m}^{*}(\frac{p_{15,m}^{*}+p_{6,m}^{*}+p_{37,m}^{*}}{p_{15,m}^{*}+p_{6,m}^{*}})\right)$.  The 2.5th and 97.5th percentiles of these draws are taken to yield a non-FPC-adjusted credible interval, denoted $\left(LL_{unajd},UL_{unadj}\right)$. FPC adjustments are implemented through a shift and scale approach, in which each posterior draw $\widehat{N}_{5,m}^{*}$ is scaled by $a=\text{min}\left(\sqrt{\frac{\widehat{Var}(\widehat{N}_5)_{FPCj}}{\widehat{Var}(\widehat{N}_5)_{unadj}}},1\right)$ (where $j=1$ for the FPC1 adjustment, and $j=2$ for the FPC2 adjustment) and shifted by $b=\widehat{N}_5(1-a)$.  The FPC-adjusted adapted Bayesian credible interval is then taken as the 2.5th and 97.5th percentiles of the shift and scaled draws, denoted $(LL_{adj}, UL_{adj})$. This approach is conceptually similar to what was proposed in \cite{Lyles2022, Lyles2023} in the 7-cell case but is here tailored explicitly for the 5-cell setting.

\section{Results}

\subsection{Simulation Study}

We conducted a number of simulations to examine the FPC adjustments for inference and to also compare competing CRC estimators under a variety of different scenarios.  In the 
\noindent first set of simulations, we examined 12 different scenarios, corresponding to three different values of $\psi$ (the known sampling probability for the anchor stream) (0.1, 0.25, 0.5), and four different values of $N$ $(500, 1000, 2500, 5000)$.  For all 12 scenarios, $N_{tot}=10,000$.  In these simulations, confidence interval coverage was evaluated according to whether the interval for that replication contained $N$ or not.  For infected individuals, "symptomatic" status was generated as $Bernoulli(0.6)$. For non-infected individuals, this was generated as $Bernoulli(0.1)$.  A non-representative sampling scheme was then implemented for Stream 1 by having 50\% and 20\% selection probabilities for symptomatic and non-symptomatic individuals, respectively. We emphasize that although we know and need to specify the nature of the sampling in Stream 1 for the purpose of simulation, we assume nothing about it for the purposes of estimation and inference. The anchor stream was implemented as a simple random sample without replacement, where the total anchor stream sample size $N_{tot}\times \psi$ was constant across replications for a given $\psi$ setting.
10,000 replications were used for each scenario.  For adapted Bayesian credible intervals, 10,000 posterior draws were used per replication.  Intervals were truncated from below at $n_c=n_2+n_4+n_6$ (the number of individuals known to be diseased from the study) and above at $N_{tot}-n_{15}$ (the number known not to be diseased). In the case that $n_6=0$, $\widehat{p}^{*}$ in $\widehat{Var}(\widehat{p}^{*})_{FPC}$ was replaced with $\widehat{p}^{*}=\frac{n_6+0.5}{n_{15}+n_{6}+1}$.  In the rare instance where $n_6=0$ and $n_{15}=0$ or $n_6+n_{15}=1$, $\widehat{Var}(\widehat{N}_5)_{FPC1}$ and $\widehat{Var}(\widehat{N}_5)_{FPC2}$ were replaced with $\widehat{Var}(\widehat{N}_5)_{unadj}$.

In the second, third, and fourth sets of simulations, $N_{tot}$ was lowered to $1000, 500$ and $250$ respectively to assess the performance of these methods in smaller population settings.  In the fifth set of simulations, we set the simulation population to be similar to the CRISP cohort with $N_{tot}=1029$, $\psi=0.194$ (to yield a random sample of size 200) and $N=156$ (the analysis of \cite{Ge2023} gave a point estimate of $\widehat{N}=156.2$ for the CRISP cohort).  We then varied the proportion of diseased individuals that were symptomatic ($p_{symp|flu}$) and the probability of being sampled into Stream 1 given symptomatic status ($p_{1|symp}$) in order to assess the performance of these methods in populations similar to the CRISP cohort under different simulation settings of the unrepresentative sampling mechanisms for Stream 1. Again, we emphasize that although we need to specify the unrepresentative sampling mechanisms of Stream 1 for the purposes of simulation, we assume nothing about it for estimation and inference.

Table 5 contains the results of the first set of simulations.  All three point estimators of $N$ are approximately unbiased in each scenario.  For $\widehat{N}_5$, it can be seen that the FPC1 adjustment to the standard error (equation 8) matches the empirical standard deviation remarkably well.  Meanwhile, the FPC2-adjusted and the unadjusted standard errors (equations 10 and 6) are generally too large, as becomes even more apparent for larger values of $\psi$ (although the FPC2-adjusted standard error is generally more accurate than the unadjusted).


\newpage
\clearpage
\begin{sidewaystable}[p]
\fontsize{8pt}{8pt}\selectfont
\caption{Simulations to Compare Different Estimators of $N$, With $N_{tot}=10,000$}
\begin{center}
\begin{threeparttable}
\begin{tabular}{|c|c|c|c|c|c|c|}
\hline
& \multicolumn{2}{|c|}{$\psi=0.1$} & \multicolumn{2}{|c|}{$\psi=0.25$} & \multicolumn{2}{|c|}{$\psi=0.5$} \\
\hline
Estimator  & \makecell{Mean (SD) \\ {[avg. SE]}} & \makecell{CI coverage \\ {[avg. width]}} & \makecell{Mean (SD) \\ {[avg. SE]}} & \makecell{CI coverage \\ {[avg. width]}} & \makecell{Mean (SD) \\ {[avg. SE]}} & \makecell{CI coverage \\ {[avg. width]}} \\
\hline
\multicolumn{7}{|c|}{$N=500$}\\
\hline
$\hat{N}_5$ \tnote{[a]} & \makecell{499.9 (52.0) \\ {[56.0]}, {[51.7]}, {[53.4]}} & \begin{tabular}{ccc}
0.962 & 0.945 & 0.952 \\
{[219.7]} & {[202.5]} & {[209.1]} \\
0.966 & 0.949 &0.956 \\
{[220.4]} & {[203.2]} & {[209.9]} 
\end{tabular} & \makecell{500.3 (30.1) \\ {[37.0]}, {[29.9]}, {[32.7]}} & \begin{tabular}{ccc}
0.983 & 0.948 & 0.964 \\
{[145.1]} & {[117.3]} & {[128.3]} \\
0.981 & 0.946 &0.967 \\
{[145.2]} & {[117.4]} & {[128.5]} 
\end{tabular} & \makecell{499.7 (17.3) \\ {[27.8]}, {[17.3]}, {[21.8]}} & \begin{tabular}{ccc}
0.999 & 0.946 & 0.985 \\
{[109.0]} & {[67.7]} & {[85.4]} \\
0.999 & 0.947 &0.987 \\
{[109.1]} & {[67.8]} & {[85.5]} 
\end{tabular} \\
\hline
$\hat{N}_{RS}$ \tnote{[b]} & \makecell{499.4 (65.4) \\ {[68.7]}, {[65.2]}} & \begin{tabular}{cc}
0.955 & 0.950 \\
{[269.4]} & {[255.7]} \\
 & 0.952 \\
 & {[256.1]}
\end{tabular} & \makecell{500.3 (38.3) \\ {[43.6]}, {[37.7]}} & \begin{tabular}{cc}
0.973 & 0.945 \\
{[170.8]} & {[147.9]} \\
 & 0.947 \\
 & {[148.0]}
\end{tabular} & \makecell{499.8 (21.7) \\ {[30.8]}, {[21.8]}} & \begin{tabular}{cc}
0.994 & 0.951 \\
{[120.8]} & {[85.4]} \\
 & 0.951 \\
 & {[85.4]}
\end{tabular} \\
\hline
$\hat{N}_{Chap}$ \tnote{[c]} & \makecell{500.6 (88.8) \\ {[84.1]}} & \makecell{0.956 \\ {[380.1]}} & \makecell{500.2 (50.0) \\ {[48.8]}} & \makecell{0.953 \\ {[203.5]}} & \makecell{499.4 (28.5) \\ {[28.2]}} & \makecell{0.950 \\ {[115.0]}} \\
\hline
\multicolumn{7}{|c|}{$N=1000$}\\
\hline
$\hat{N}_5$ \tnote{[a]} & \makecell{1001.9 (72.9) \\ {[78.1]}, {[72.1]}, {[74.3]}} & \begin{tabular}{ccc}
0.960 & 0.945 & 0.951 \\
{[306.1]} & {[282.7]} & {[291.4]} \\
0.963 & 0.945 &0.952 \\
{[306.2]} & {[282.9]} & {[291.5]} 
\end{tabular} & \makecell{1000.0 (41.5) \\ {[51.3]}, {[41.6]}, {[45.3]}} & \begin{tabular}{ccc}
0.984 & 0.950 & 0.968 \\
{[201.1]} & {[163.2]} & {[177.7]} \\
0.985 & 0.950 &0.967 \\
{[201.1]} & {[163.2]} & {[177.7]} 
\end{tabular} & \makecell{1000.1 (24.1) \\ {[38.4]}, {[24.0]}, {[30.0]}} & \begin{tabular}{ccc}
0.998 & 0.950 & 0.985 \\
{[150.7]} & {[94.2]} & {[117.6]} \\
0.998 & 0.950 &0.985 \\
{[150.7]} & {[94.3]} & {[117.6]} 
\end{tabular} \\
\hline
$\hat{N}_{RS}$ \tnote{[b]}  & \makecell{1001.7 (90.5) \\ {[94.8]}, {[90.0]}} & \begin{tabular}{cc}
0.964 & 0.952 \\
{[371.7]} & {[352.8]} \\
 & 0.947 \\
 & {[352.8]}
\end{tabular} & \makecell{1000.3 (52.3) \\ {[60.0]}, {[52.0]}} & \begin{tabular}{cc}
0.973 & 0.948 \\
{[235.1]} & {[203.7]} \\
 & 0.948 \\
 & {[203.7]}
\end{tabular} & \makecell{999.9 (29.9) \\ {[42.4]}, {[30.0]}} & \begin{tabular}{cc}
0.994 & 0.950 \\
{[166.3]} & {[117.6]} \\
 & 0.952 \\
 & {[117.6]}
\end{tabular} \\
\hline
$\hat{N}_{Chap}$ \tnote{[c]} & \makecell{1002.8 (125.4) \\ {[120.2]}} & \makecell{0.950 \\ {[505.1]}} & \makecell{999.5 (69.9) \\ {[69.4]}} & \makecell{0.953 \\ {[280.3]}} & \makecell{1000.5 (41.1) \\ {[40.2]}} & \makecell{0.949 \\ {[160.6]}} \\
\hline
\multicolumn{7}{|c|}{$N=2500$}\\
\hline
$\hat{N}_5$ \tnote{[a]} & \makecell{2499.6 (107.1) \\ {[115.7]}, {[107.3]}, {[110.1]}} & \begin{tabular}{ccc}
0.965 & 0.950 & 0.954 \\
{[453.4]} & {[420.7]} & {[431.4]} \\
0.966 & 0.951 &0.957 \\
{[452.9]} & {[420.2]} & {[431.0]} 
\end{tabular} & \makecell{2498.9 (61.1) \\ {[75.6]}, {[62.0]}, {[66.6]}} & \begin{tabular}{ccc}
0.985 & 0.952 & 0.967 \\
{[296.2]} & {[242.9]} & {[261.0]} \\
0.985 & 0.953 &0.967 \\
{[296.0]} & {[242.7]} & {[260.8]} 
\end{tabular} & \makecell{2500.7 (35.4) \\ {[56.2]}, {[35.8]}, {[43.3]}} & \begin{tabular}{ccc}
0.999 & 0.954 & 0.984 \\
{[220.2]} & {[140.3]} & {[169.8]} \\
0.999 & 0.953 &0.984 \\
{[220.1]} & {[140.2]} & {[169.7]} 
\end{tabular} \\
\hline
$\hat{N}_{RS}$ \tnote{[b]} & 
 \makecell{2498.2 (130.0) \\ {[136.8]}, {[129.9]}} & \begin{tabular}{cc}
0.958 & 0.949 \\
{[536.3]} & {[509.0]} \\
 & 0.951 \\
 & {[508.5]}
\end{tabular} & \makecell{2498.6 (74.7) \\ {[86.6]}, {[75.0]}} & \begin{tabular}{cc}
0.977 & 0.954 \\
{[339.3]} & {[293.9]} \\
 & 0.951 \\
 & {[293.8]}
\end{tabular} & \makecell{2500.6 (42.9) \\ {[61.2]}, {[43.3]}} & \begin{tabular}{cc}
0.995 & 0.949 \\
{[240.0]} & {[169.8]} \\
 & 0.951 \\
 & {[169.7]}
\end{tabular} \\
\hline
$\hat{N}_{Chap}$ \tnote{[c]} & \makecell{2502.9 (194.4) \\ {[190.7]}} & \makecell{0.951 \\ {[768.3]}} & \makecell{2499.7 (109.7) \\ {[110.0]}} & \makecell{0.952 \\ {[436.2]}} & \makecell{2500.9 (63.5) \\ {[63.5]}} & \makecell{0.951 \\ {[251.1]}} \\
\hline
\multicolumn{7}{|c|}{$N=5000$}\\
\hline
$\hat{N}_5$ \tnote{[a]} & \makecell{5002.5 (133.3) \\ {[140.3]}, {[131.2]}, {[133.4]}} & \begin{tabular}{ccc}
0.960 & 0.944 & 0.948 \\
{[550.1]} & {[514.4]} & {[523.1]} \\
0.960 & 0.944 &0.948 \\
{[549.1]} & {[513.4]} & {[522.0]} 
\end{tabular} & \makecell{5000.1 (74.7) \\ {[90.7]}, {[75.7]}, {[79.5]}} & \begin{tabular}{ccc}
0.984 & 0.953 & 0.963 \\
{[355.7]} & {[296.9]} & {[311.8]} \\
0.983 & 0.954 &0.963 \\
{[355.3]} & {[296.6]} & {[311.4]} 
\end{tabular} & \makecell{4999.8 (43.8) \\ {[66.4]}, {[43.7]}, {[50.0]}} & \begin{tabular}{ccc}
0.997 & 0.952 & 0.976 \\
{[260.4]} & {[171.4]} & {[196.0]} \\
0.997 & 0.952 &0.976 \\
{[260.2]} & {[171.3]} & {[195.8]} 
\end{tabular} \\
\hline
$\hat{N}_{RS}$ \tnote{[b]} & \makecell{5001.7 (151.7) \\ {[158.0]}, {[150.0]}} & \begin{tabular}{cc}
0.956 & 0.950 \\
{[619.5]} & {[588.0]} \\
 & 0.950 \\
 & {[587.3]}
\end{tabular} & \makecell{5000.2 (85.6) \\ {[100.0]}, {[86.6]}} & \begin{tabular}{cc}
0.977 & 0.954 \\
{[391.9]} & {[339.5]} \\
 & 0.954 \\
 & {[339.3]}
\end{tabular} & \makecell{4999.7 (50.0) \\ {[70.7]}, {[50.0]}} & \begin{tabular}{cc}
0.994 & 0.951 \\
{[277.2]} & {[196.0]} \\
 & 0.956 \\
 & {[196.0]}
\end{tabular} \\
\hline
$\hat{N}_{Chap}$ \tnote{[c]} & \makecell{5005.5 (275.7) \\ {[270.7]}} & \makecell{0.948 \\ {[1075.7]}} & \makecell{5000.3 (157.8) \\ {[156.1]}} & \makecell{0.947 \\ {[615.6]}} & \makecell{5000.2 (90.1) \\ {[90.2]}} & \makecell{0.950 \\ {[354.9]}} \\
\hline
\end{tabular}
\begin{tablenotes}
\item[a] Reported standard errors are the unadjusted (left), the FPC1-adjusted (center), and the FPC2-adjusted (right); Confidence intervals are a Wald unadjusted (top left), Wald FPC1-adjusted (top center) Wald-FPC2 adjusted (top right), unadjusted Bayesian credible interval (bottom left), FPC1-adjusted Bayesian credible interval (bottom center), and FPC2-adjusted Bayesian credible interval (bottom right).
\item[b] Reported standard errors are the non-FPC-adjusted (left) and Cochran's FPC-adjusted (right); Confidence intervals are non-FPC-adjusted Wald (top left), Cochran's FPC-adjusted Wald (top right), and a FPC-adjusted Jeffreys prior-based credible interval (bottom right) (\cite{Lyles2023}).

\item[c] Reported confidence interval is a logit-based confidence interval given by \cite{Sadinle2009}.
\end{tablenotes}
\end{threeparttable}
\end{center}
\end{sidewaystable}

\clearpage
\newpage


\newpage
\clearpage
\begin{sidewaystable}[p]
\fontsize{8pt}{8pt}\selectfont
\caption{Simulations to Compare Different Estimators of $N$, With $N_{tot}=1000$}
\begin{center}
\begin{threeparttable}
\begin{tabular}{|c|c|c|c|c|c|c|}
\hline
& \multicolumn{2}{|c|}{$\psi=0.1$} & \multicolumn{2}{|c|}{$\psi=0.25$} & \multicolumn{2}{|c|}{$\psi=0.5$} \\
\hline
Estimator & \makecell{Mean (SD) \\ {[avg. SE]}} & \makecell{CI coverage \\ {[avg. width]}} & \makecell{Mean (SD) \\ {[avg. SE]}} & \makecell{CI coverage \\ {[avg. width]}} & \makecell{Mean (SD) \\ {[avg. SE]}} & \makecell{CI coverage \\ {[avg. width]}} \\
\hline
\multicolumn{7}{|c|}{$N=50$}\\
\hline
$\hat{N}_5$ \tnote{[a]} &  \makecell{50.2 (16.4) \\ {[17.4]}, {[15.9]}, {[16.5]}} & \begin{tabular}{ccc}
0.920 & 0.872 & 0.903 \\
{[61.0]} & {[57.4]} & {[58.8]} \\
0.954 & 0.936 &0.945 \\
{[70.1]} & {[64.2]} & {[66.5]} 
\end{tabular} & \makecell{50.1 (9.4) \\ {[11.6]}, {[9.3]}, {[10.3]}} & \begin{tabular}{ccc}
0.969 & 0.930 & 0.954 \\
{[43.9]} & {[36.4]} & {[39.5]} \\
0.985 & 0.942 &0.966 \\
{[46.1]} & {[37.3]} & {[41.0]} 
\end{tabular} & \makecell{50.1 (5.5) \\ {[8.8]}, {[5.4]}, {[6.9]}} & \begin{tabular}{ccc}
0.994 & 0.944 & 0.982 \\
{[32.3]} & {[21.3]} & {[26.7]} \\
0.998 & 0.947 &0.984 \\
{[34.4]} & {[21.5]} & {[27.3]} 
\end{tabular} \\
\hline
$\hat{N}_{RS}$ \tnote{[b]} &  \makecell{50.0 (20.7) \\ {[21.2]}, {[20.3]}} & \begin{tabular}{cc}
0.897 & 0.891 \\
{[71.0]} & {[68.8]} \\
 & 0.951 \\
 & {[76.3]}
\end{tabular} & \makecell{50.0 (11.8) \\ {[13.7]}, {[11.9]}} & \begin{tabular}{cc}
0.961 & 0.951 \\
{[48.3]} & {[43.3]} \\
 & 0.957 \\
 & {[45.4]}
\end{tabular} & \makecell{50.0 (6.9) \\ {[9.7]}, {[6.9]}} & \begin{tabular}{cc}
0.986 & 0.932 \\
{[34.0]} & {[25.9]} \\
 & 0.942 \\
 & {[26.5]}
\end{tabular} \\
\hline
$\hat{N}_{Chap}$ \tnote{[c]}  & \makecell{46.3 (22.4) \\ {[18.8]}} & \makecell{0.996 \\ {[270.1]}} & \makecell{50.1 (18.1) \\ {[14.5]}} & \makecell{0.976 \\ {[112.1]}} & \makecell{50.2 (9.9) \\ {[8.6]}} & \makecell{0.958 \\ {[50.5]}} \\
\hline
\multicolumn{7}{|c|}{$N=100$}\\
\hline
$\hat{N}_5$ \tnote{[a]} & \makecell{100.2 (22.6) \\ {[23.9]}, {[22.1]}, {[22.9]}} & \begin{tabular}{ccc}
0.940 & 0.920 & 0.933 \\
{[91.9]} & {[85.6]} & {[88.2]} \\
0.963 & 0.941 &0.954 \\
{[95.0]} & {[87.7]} & {[90.9]} 
\end{tabular} & \makecell{100.0 (12.9) \\ {[16.0]}, {[12.9]}, {[14.2]}} & \begin{tabular}{ccc}
0.979 & 0.942 & 0.965 \\
{[62.5]} & {[50.5]} & {[55.5]} \\
0.986 & 0.950 &0.970 \\
{[62.9]} & {[50.7]} & {[55.8]} 
\end{tabular} & \makecell{100.0 (7.5) \\ {[12.1]}, {[7.5]}, {[9.5]}} & \begin{tabular}{ccc}
0.998 & 0.944 & 0.985 \\
{[47.2]} & {[29.2]} & {[37.2]} \\
0.999 & 0.948 &0.987 \\
{[47.4]} & {[29.3]} & {[37.3]} 
\end{tabular} \\
\hline
$\hat{N}_{RS}$ \tnote{[b]} &  \makecell{100.3 (28.7) \\ {[29.6]}, {[28.3]}} & \begin{tabular}{cc}
0.941 & 0.941 \\
{[106.3]} & {[102.5]} \\
 & 0.948 \\
 & {[107.1]}
\end{tabular} & \makecell{100.0 (16.3) \\ {[18.9]}, {[16.4]}} & \begin{tabular}{cc}
0.965 & 0.937 \\
{[72.2]} & {[63.4]} \\
 & 0.951 \\
 & {[63.8]}
\end{tabular} & \makecell{100.0 (9.5) \\ {[13.4]}, {[9.5]}} & \begin{tabular}{cc}
0.991 & 0.951 \\
{[51.3]} & {[37.0]} \\
 & 0.957 \\
 & {[37.1]}
\end{tabular} \\
\hline
$\hat{N}_{Chap}$ \tnote{[c]} &  \makecell{99.3 (42.0) \\ {[33.1]}} & \makecell{0.984 \\ {[251.7]}} & \makecell{100.2 (23.1) \\ {[20.5]}} & \makecell{0.958 \\ {[109.7]}} & \makecell{100.0 (12.5) \\ {[11.9]}} & \makecell{0.956 \\ {[56.5]}} \\
\hline
\multicolumn{7}{|c|}{$N=250$}\\
\hline
$\hat{N}_5$ \tnote{[a]} & \makecell{250.2 (33.8) \\ {[36.2]}, {[33.7]}, {[34.6]}} & \begin{tabular}{ccc}
0.955 & 0.942 & 0.948 \\
{[141.8]} & {[132.1]} & {[135.6]} \\
0.964 & 0.949 &0.955 \\
{[140.8]} & {[131.1]} & {[134.6]} 
\end{tabular} & \makecell{250.3 (19.6) \\ {[23.8]}, {[19.5]}, {[21.0]}} & \begin{tabular}{ccc}
0.980 & 0.947 & 0.963 \\
{[93.3]} & {[76.5]} & {[82.4]} \\
0.982 & 0.947 &0.965 \\
{[93.0]} & {[76.2]} & {[82.1]} 
\end{tabular} & \makecell{250.1 (11.2) \\ {[17.7]}, {[11.3]}, {[13.7]}} & \begin{tabular}{ccc}
0.998 & 0.953 & 0.985 \\
{[69.5]} & {[44.2]} & {[53.7]} \\
0.998 & 0.952 &0.984 \\
{[69.4]} & {[44.1]} & {[53.6]} 
\end{tabular} \\
\hline
$\hat{N}_{RS}$ \tnote{[b]} &  \makecell{250.1 (41.3) \\ {[43.0]}, {[41.0]}} & \begin{tabular}{cc}
0.958 & 0.946 \\
{[168.0]} & {[160.3]} \\
 & 0.946 \\
 & {[159.1]}
\end{tabular} & \makecell{250.3 (23.9) \\ {[27.3]}, {[23.7]}} & \begin{tabular}{cc}
0.969 & 0.945 \\
{[107.2]} & {[93.0]} \\
 & 0.955 \\
 & {[92.6]}
\end{tabular} & \makecell{250.2 (13.7) \\ {[19.4]}, {[13.7]}} & \begin{tabular}{cc}
0.992 & 0.945 \\
{[75.9]} & {[53.7]} \\
 & 0.954 \\
 & {[53.6]}
\end{tabular} \\
\hline
$\hat{N}_{Chap}$ \tnote{[c]} & \makecell{250.4 (65.0) \\ {[57.8]}} & \makecell{0.963 \\ {[292.8]}} & \makecell{250.2 (35.0) \\ {[33.9]}} & \makecell{0.956 \\ {[149.9]}} & \makecell{249.9 (20.0) \\ {[19.7]}} & \makecell{0.955 \\ {[83.3]}} \\
\hline
\multicolumn{7}{|c|}{$N=500$}\\
\hline
$\hat{N}_5$ \tnote{[a]} & \makecell{500.5 (41.5) \\ {[44.2]}, {[41.5]}, {[42.2]}} & \begin{tabular}{ccc}
0.959 & 0.946 & 0.950 \\
{[173.1]} & {[162.7]} & {[165.5]} \\
0.963 & 0.950 &0.954 \\
{[170.6]} & {[160.4]} & {[163.1]} 
\end{tabular} & \makecell{500.0 (24.2) \\ {[28.6]}, {[23.9]}, {[25.2]}} & \begin{tabular}{ccc}
0.981 & 0.943 & 0.957 \\
{[112.3]} & {[93.9]} & {[98.6]} \\
0.982 & 0.946 &0.958 \\
{[111.6]} & {[93.3]} & {[98.0]} 
\end{tabular} & \makecell{500.2 (13.7) \\ {[21.0]}, {[13.8]}, {[15.8]}} & \begin{tabular}{ccc}
0.997 & 0.952 & 0.975 \\
{[82.3]} & {[54.2]} & {[62.0]} \\
0.997 & 0.951 &0.976 \\
{[82.0]} & {[54.0]} & {[61.8]} 
\end{tabular} \\
\hline
$\hat{N}_{RS}$ \tnote{[b]} & \makecell{500.5 (47.3) \\ {[49.8]}, {[47.5]}} & \begin{tabular}{cc}
0.954 & 0.954 \\
{[195.1]} & {[186.0]} \\
 & 0.954 \\
 & {[183.8]}
\end{tabular} & \makecell{500.0 (27.7) \\ {[31.6]}, {[27.4]}} & \begin{tabular}{cc}
0.977 & 0.949 \\
{[123.8]} & {[107.4]} \\
 & 0.949 \\
 & {[106.9]}
\end{tabular} & \makecell{500.2 (15.6) \\ {[22.3]}, {[15.8]}} & \begin{tabular}{cc}
0.993 & 0.952 \\
{[87.6]} & {[62.0]} \\
 & 0.952 \\
 & {[61.9]}
\end{tabular} \\
\hline
$\hat{N}_{Chap}$ \tnote{[c]} &  \makecell{500.0 (88.1) \\ {[83.8]}} & \makecell{0.956 \\ {[357.3]}} & \makecell{499.9 (50.2) \\ {[48.9]}} & \makecell{0.951 \\ {[203.3]}} & \makecell{499.9 (28.6) \\ {[28.3]}} & \makecell{0.952 \\ {[115.4]}} \\
\hline
\end{tabular} 
\begin{tablenotes}
\item[a] Reported standard errors are the unadjusted (left), the FPC1-adjusted (center), and the FPC2-adjusted (right); Confidence intervals are a Wald unadjusted (top left), Wald FPC1-adjusted (top center) Wald-FPC2 adjusted (top right), unadjusted Bayesian credible interval (bottom left), FPC1-adjusted Bayesian credible interval (bottom center), and FPC2-adjusted Bayesian credible interval (bottom right).
\item[b] Reported standard errors are the non-FPC-adjusted (left) and Cochran's FPC-adjusted (right); Confidence intervals are non-FPC-adjusted Wald (top left), Cochran's FPC-adjusted Wald (top right), and a FPC-adjusted Jeffreys prior-based credible interval (bottom right) (\cite{Lyles2023}).

\item[c] Reported confidence interval is a logit-based confidence interval given by \cite{Sadinle2009}.
\end{tablenotes}
\end{threeparttable}
\end{center}
\end{sidewaystable}
\clearpage

 Accordingly, Wald and adapted Bayesian credible intervals based on the FPC2-adjusted and unadjusted variances are often too conservative, particularly at larger values of $\psi$ (again, we note that the FPC2-adjusted intervals are more precise than the unadjusted ones though). Meanwhile, Wald and adjusted Bayesian credible interval coverages based on FPC1 are typically quite close to the nominal 95\% level, and the corresponding widths are much narrower than those using the unadjusted variance.   When comparing Wald-type intervals vs adjusted Bayesian credible intervals for the $\widehat{N}_5$ estimator, we find that the latter more consistently achieve near-nominal 95\% coverage.

When comparing estimators, we see that $\widehat{N}_5$ generally performs better than $\widehat{N}_{RS}$ and $\widehat{N}_{Chap}$, offering considerably more precision.  This is expected, as $\widehat{N}_5$ is the MLE based on the full data.

In the 2nd, 3rd, and 4th sets of simulations (Tables 2, B1, and B2 respectively), our findings are largely similar to what we found for the first set, with the FPC1-adjusted standard error generally matching the empirical standard deviation more accurately than the FPC2 and unadjusted standard errors.  When $N_{tot}=250$ and $N_{tot}=500$, we do find some cases of undercoverage for the FPC1-adjusted Bayesian credible interval, though in most cases it achieves near the 95\% nominal level and provides more precision in inferences.  We also find that Chapman's estimator displays downward bias when $N$ and $\psi$ are smaller.  This is in keeping with the finding in  \cite{Wittes1972} that $\widehat{N}_{Chap}$ displays downward bias when $\frac{n_{1.}n_{.1}}{N}$ is small, where $n_{1.}$ is the number of the diseased captured by Stream 1, and $n_{.1}$ is the number of the diseased captured by Stream 2.  To see this, observe that $E(\frac{n_{1.}n_{.1}}{N})\propto\frac{N(\psi)(N)}{N}=N\psi$, which is small when both $N$ and $\psi$ are small.

In the fifth set of simulations (Table B3), we found that all three point estimators of $N$ were generally unbiased for populations similar to the CRISP cohort.  Again, we found that the FPC1-adjusted standard error generally matched the empirical standard deviation more accurately than the FPC2 and unadjusted standard errors and that adjusted Bayesian credible interval approaches more consistently achieved the 95\% coverage level as compared to Wald intervals.  Moreover, $\widehat{N}_5$ again offered more precision than $\widehat{N}_{RS}$ and $\widehat{N}_{Chap}$.

Based on these results, we would generally recommend the FPC1-adjusted standard error and Bayesian credible interval to accompany $\widehat{N}_5$ rather than the FPC2-adjusted or unadjusted standard errors and corresponding intervals.

\subsection{Real data analysis: CRISP Cancer Recurrence}

The CRISP (Cancer Recurrence Information and Surveillance Program) was established to monitor cancer recurrences in the Georgia Cancer Registry (GCR).  The program monitors recurrences for four cancer types (breast, colorectal, lymphoma, and prostate) using up to six data streams for each cancer type to signal a potential recurrence (\cite{Ge2023}).  We seek to estimate the number of breast cancer recurrences among registry patients initially diagnosed with breast cancer during 2013-2015 who had their initial surgery performed at one of the five hospitals with Commission on Cancer accreditation (CoC) in the five metro Atlanta counties.  Since interest is solely in the number of recurrences in this fixed, finite population, FPC adjustments would be necessary.  In this example, we seek to illustrate the precision gains that can be obtained by synthesizing information from the anchor and non-anchor streams and by implementing our proposed FPC corrections.


\begin{table}[H]
\centering
\caption{Cell counts for the CRISP data.}
\begin{tabular}{|>{$}l<{$} p{9cm} r|}
\hline
\text{Cell} & \text{Cell Type} & \text{Cell Count} \\
\hline
n_{15} & Sampled in both streams, tested negative \newline OR Sampled in Stream 2 but not Stream 1, tested negative. & 169 \\
\hline
n_{2}  & Sampled in both streams, tested positive. & 12 \\
\hline
n_{4}  & Sampled in Stream 1 but not 2, tested positive. & 52 \\
\hline
n_{6}  & Sampled in Stream 2 but not 1, tested positive. & 19 \\
\hline
n_{37} & Not sampled in either stream \newline OR Sampled in Stream 1 but not 2, tested negative. & 777 \\
\hline
N_{\text{tot}} & & 1029 \\
\hline
\end{tabular}

\end{table}

The finite, closed target population consists of $N_{tot}=1029$ patients.  Hospitals participating in the CoC program report cancer recurrences among patients diagnosed in their facilities.  This initial data stream (which we refer to as the CoC data stream) has the potential for misclassification (false positives).  However, of these recurrences, all but 10 were selected for validation, yielding what we refer to as Stream 1 (or as the CoC-validated data stream).  On May 1, 2022, an anchor stream random sample was drawn independently of Stream 1.  This sample consisted of 200 patients, of which 31 were identified as having a breast cancer recurrence.  These recurrences were identified via CRISP protocol-based medical report abstraction, which we treat as a gold-standard test for the purposes of this analysis.  The same protocol was used to validate the CoC-identified cases.

Table 7 summarizes the capture profiles of the 1029 patients.  Of 83 signaled recurrences from the CoC data stream, 73 were selected for validation, and of these 67 were validated as true recurrences. 3 CoC-validated recurrences that were judged as non-cases in the anchor stream were placed in cell $n_{15}$. \cite{Ge2023} previously analyzed these data, but instead of using the CoC-validated data for Stream 1, they used the unvalidated CoC recurrence cases to illustrate an approach that accounts for misclassification. This analysis yielded an estimate of 156.2 breast cancer recurrences (95\% Wald CI=[115.7, 201.9], 95\% Unadjusted Bayesian credible interval=[118.5, 198.8]). Here, we are using the CoC-validated data so that adjustments for false positives are not necessary.

Table 8 contains the results of competing estimation and inference procedures for the number of breast cancer recurrences.  The point estimators generally agree, but the associated intervals differ appreciably.  The estimated number of recurrences based on using only the anchor stream is $\widehat{N}_{RS}=159.5$ (95\% CI=[113.1, 205.9]).  This interval is wider than those associated with inferences using the anchor stream design based 5-cell estimator, as the random sample estimator integrates no information from Stream 1.  The Chapman estimate is $\widehat{N}_{Chap}=159.0$ (95\% CI=[119.7, 263.0]) with an associated confidence interval (\cite{Sadinle2009}) that is quite wide.

Confidence intervals associated with $\widehat{N}_5$ are narrower than those associated with $\widehat{N}_{RS}$, illustrating the precision gains drawn from incorporating Stream 1.  Additionally, FPC-adjusted intervals associated with $\widehat{N}_5$ are narrower than non-FPC-adjusted intervals, showing the improved precision due to accounting for finite population effects.  Following our recommendation from the simulation results that we should report $\widehat{N}_5$ with an FPC1-adjusted Bayesian credible interval, our reported estimate of breast cancer recurrences among metro Atlanta-based Georgia Cancer Registry Patients is 161.5 (95\% CI=[130.1, 203.5]).


\begin{threeparttable}
\centering
\caption{Point estimates and accompanying confidence intervals for the number of breast cancer recurrences in the CRISP cohort.}
\begin{tabular}{|c c c|}
\hline
Estimator & \makecell{$\hat{N}$\\{[SE]}} & CI \\
\hline
$\widehat{N}_5$ \tnote{a} & 
\makecell{161.5 \\ {[22.3], [19.1], [20.3]}} &
\begin{tabular}{ccc}
{[117.8, 205.3]} & {[124.1, 198.9]} & {[121.7, 201.3]}\\
{[124.1, 212.5]} & {[130.1, 203.5]} & {[126.7, 206.0]}\\
\end{tabular} \\
\hline
$\widehat{N}_{RS}$ \tnote{b} & 
\makecell{159.5 \\ {[26.3], [23.7]}} &
\begin{tabular}{cc}
{[107.9, 211.1]}& {[113.1, 206.0]}\\
                & {[117.8, 210.4]}\\
\end{tabular}\\
\hline
$\widehat{N}_{Chap}$ \tnote{c} & 
\makecell{159.0 \\ {[29.5]}} &
$[119.7,\ 263.0]$ \\
\hline
\end{tabular}

\begin{tablenotes}
\tiny{\item[a] Reported standard errors are the unadjusted (left), the FPC1-adjusted (center), and the FPC2-adjusted (right); Confidence intervals are a Wald unadjusted (top left), Wald FPC1-adjusted (top center) Wald-FPC2 adjusted (top right), unadjusted Bayesian credible interval (bottom left), FPC1-adjusted Bayesian credible interval (bottom center), and FPC2-adjusted Bayesian credible interval (bottom right).
\item[b] Reported standard errors are the non-FPC-adjusted (left) and Cochran's FPC-adjusted (right); Confidence intervals are non-FPC-adjusted Wald (top left), Cochran's FPC-adjusted Wald (top right), and a FPC-adjusted Jeffreys prior-based credible interval (bottom right) (\cite{Lyles2023}).

\item[c] Reported confidence interval is a logit-based confidence interval given by \cite{Sadinle2009}.
}
\end{tablenotes}
\end{threeparttable}\\

R code for the simulation study and CRISP data analysis can be found through the following GitHub link: https://github.com/mcdoerf/5-Cell-FPC-Demonstration.

\section{Discussion}

In this paper, we have presented a refinement of inferences for the ``anchor stream" design under a common situation in which Stream 1 only records positive test results, and the inferential goal is to estimate the number of disease cases in a finite, closed population. We developed two FPC-adjusted variance estimators to account for finite population effects, as well as an accompanying adapted Bayesian credible interval approach for inference. Our simulation results suggest that these FPC-adjusted inferences, particularly those using the FPC1 adjustment, can yield significantly improved precision. We applied these approaches to the CRISP data, yielding an estimate of the number of breast cancer recurrences among GCR patients in the metro Atlanta area.

Extensions and limitations of our approach should be noted.  First, the methods presented in this paper can easily be extended to the case where there are multiple non-anchor streams. In this situation, the multiple non-anchor streams can be combined into a single Stream 1, where a patient ``tests positive" if any of the streams flag them as a disease case. Inference then proceeds as outlined in this paper.  Additionally, these approaches can be used when the anchor stream is a stratified random sample, assuming the stratification variables are also available for Stream 1.  In this case (assuming the strata are mutually exclusive and exhaustive), disease totals are estimated across strata and then summed to estimate the case totals for the entire cohort. Variance estimates are obtained as the sum of variance estimates across strata.  Next, a limitation of our approach is that we assume Stream 1 diagnoses patients without misclassification. In some situations, this can be quite unrealistic. However, the results in this paper can help practitioners fully utilize the information from their data in the case where Stream 1 can be assumed accurate, such as the CRISP example.  Moreover, we believe that the results presented here provide an informative example of the importance of implementing FPC adjustments in non-standard sampling designs.  For a discussion of the anchor stream design with misclassification, see \cite{Ge2023, Ge2024}.

Our results illustrate the appeal and practicality of the anchor stream design.  Without an anchor stream, the arbitrarily non-representative Stream 1 yields biased estimates and cannot be used for valid inference on its own. However, the incorporation of an independent anchor stream can allow for the information from Stream 1 to be utilized, yielding estimators that are more efficient than those based on the anchor stream sample alone. This design can be implemented so long as it is possible to draw a random sample from the population of interest that is independent of the biased data stream(s).  This approach can be utilized for estimation of cumulative disease cases, and more generally, estimation of the total number of individuals that have a characteristic of interest in a closed cohort setting.  We hope that our work here, in addition to providing techinical refinement of inferences, can help spread awareness of the anchor stream design and further facilitate its use in practice. 

\newpage

\printbibliography

\newpage

\section{Appendix}

\renewcommand{\theequation}{A.\arabic{equation}}
\setcounter{equation}{0}

\renewcommand{\thetable}{A\arabic{table}}
\setcounter{table}{0}

\subsection{Appendix A: Details for Derivation of FPC1 and FPC2 Adjustments}.

The FPC1 and FPC2-adjusted variances for $\widehat{N}_5$ (equations 9 and 10 respectively) were derived based on the following reparameterization of the multinomial likelihood in Table 4.

\begin{table}[ht]
\centering
\caption{5-cell case with population-level multinomial likelihood contributions.}
\label{tab:5cell_likelihood_wrapped}
\begin{tabular}{p{1.5cm} p{6cm} p{6cm}}  
\toprule
\textbf{Cell Count} & \textbf{Observation Type} & \textbf{Likelihood Contribution} \\
\midrule
$n_{15}$ & 
\parbox[t]{6cm}{Sampled in both streams, tested negative \textbf{OR} sampled in Stream 2 but not Stream 1, tested negative.} & 
\parbox[t]{6cm}{\( p_{15} =(1-\phi_{R})(1-\pi_{\overline{R}1})(\psi_{R}) \)} \\

\hline

$n_2$ & 
\parbox[t]{6cm}{Sampled in both streams, tested positive.} & 
\parbox[t]{6cm}{\( p_2 = \psi_{R} \phi_{R} \)} \\

\hline

$n_4$ & 
\parbox[t]{6cm}{Sampled in Stream 1 but not 2, tested positive.} & 
\parbox[t]{6cm}{\( p_4 = (1 - \psi_{R}) \phi_{R} \)} \\

\hline

$n_6$ & 
\parbox[t]{6cm}{Sampled in Stream 2 but not 1, tested positive.} & 
\parbox[t]{6cm}{\( p_6 = (1-\phi_{R})(\pi_{\overline{R}1})(\psi_{R}) \)} \\

\hline

$n_{37}$ & 
\parbox[t]{6cm}{Not sampled in either stream \textbf{OR} sampled in Stream 1 but not 2, tested negative.} & 
\parbox[t]{6cm}{\( p_{37} = (1-\phi_{R})(1-\psi_{R}) \)} \\
\bottomrule
\end{tabular}
\end{table}

\noindent where $\phi_{R}=\pi_{S1}\phi$, $\psi_{R}=\psi$, and $\pi_{\overline{R}1}=\frac{\pi_{\overline{S}1}(1-\phi)}{1-\pi_{S1}\phi}$.  To interpret these parameters, we first need to note that there is a difference between being \textit{sampled} and being \textit{recorded} in the 5-cell setting.  Those who are not diseased but are sampled into Stream 1 are not recorded as having been sampled.

Then, we have that 

\begin{align*}
\phi_{R}&=Pr(\text{Diseased}|\text{Sampled in Stream 1})Pr(\text{Sampled in Stream 1})\\
&=Pr(\text{Recorded in Stream 1}). \\
\psi_{R}&=Pr(\text{Sampled in Stream 2})\\
&=Pr(\text{Recorded in Stream 2}).\\
\pi_{\overline{R}1}&=\frac{Pr(\text{Diseased} | \text{Not sampled in Stream 1})Pr(\text{Not sampled in Stream 1)}}{1-Pr(\text{Diseased | Sampled in Stream 1})Pr(\text{Sampled in Stream 1})}\\
&=\frac{Pr(\text{Diseased} \cap \text{Not sampled in Stream 1})}{Pr(\text{Not Sampled in 1})+Pr(\text{Sampled in 1}\cap \text{Not Diseased})}\\
&=Pr(\text{Diseased}|\text{Not recorded in Stream 1}). \\     
\end{align*}

We then have that for the disease prevalence $\pi$,

\begin{equation}
\begin{split}
\pi&=Pr(\text{Diseased})\\
&=Pr(\text{Diseased}|\text{Not record in Stream 1})Pr(\text{Not recorded in Stream 1})\\
&+Pr(\text{Diseased}|\text{Recorded in Stream 1})Pr(\text{Recorded in Stream 1})\\
&=\pi_{\overline{R}1}(1-\phi_{R})+\pi_{R1}\phi_R,
\end{split}
\end{equation}

\noindent where $\pi_{R1}=Pr(\text{Diseased}|\text{Recorded in Stream 1})=1.$

Basic calculations then yield that the MLEs of $\pi_{\overline{R}1}$ and $\phi_{R}$ are $\widehat{\pi}_{\overline{R}1}=\frac{n_{6}}{n_{15}+n_{6}}$ and $\widehat{\phi}_{R}=\frac{n_2+n_4}{N_{tot}}$ respectively.

This then gives that the MLE of $\pi$ is,

\begin{equation}
    \widehat{\pi}_5=\widehat{\pi}_{\overline{R}1}(1-\widehat{\phi}_R)+\pi_{R1}\widehat{\phi}_R
\end{equation}

\noindent that is, a weighted average of $Pr(\text{Diseased}|\text{Not recorded in Stream 1})$ and 

\noindent $Pr(\text{Diseased} | \text{Recorded in Stream 1})$, with weight $w=\widehat{\phi}_R$.  It is straightforward to show that equations (A.2) and (3) are identical.

FPC1 is then derived by treating the weight $w$ as fixed. 

\begin{equation*}
\begin{split}
V(\widehat{\pi}_5)&=V(\widehat{\pi}_{\overline{R}1}(1-w)+\pi_{R1}(w))\\ 
&=(1-w)^2V(\widehat{\pi}_{\overline{R}1}).
\end{split}
\end{equation*}

To derive the FPC-adjusted variance of $\widehat{\pi}_{\overline{R}1}$, we follow an approach similar to \cite{Lyles2023} in which we view $\widehat{\pi}_{\overline{R}1}$ as a sample proportion of size $n_{RS}^{*}=n_{15}+n_{6}$ drawn from a finite population of size $N_{tot}^{*}=n_{15}+n_{6}+n_{37}$.  Consider the following.

\begin{equation}
    \begin{split}
 V(\widehat{\pi}_{\overline{R}1}) &=V(\frac{n_6}{n_{15}+n_6})\\
 &=E[V(\frac{n_6}{n_{15}+n_6}|n_{15}+n_6, n_{15}+n_6+n_{37})]+V[E(\frac{n_{6}}{n_{15}+n_{6}}|n_{15}+n_6, n_{15}+n_{6}+n_{37})].
    \end{split}
\end{equation}

Under the multinomial model in Table A1, we have that,

\begin{equation*}
\tiny{
    \begin{split}
        P(N_{6}=n_6|N_{15}+N_{6}=n_{15}+n_{6}, N_{15}+N_{6}+N_{37}=n_{15}+n_{6}+n_{37})&=\\
        \frac{P(N_6=n_6, N_{15}=n_{15}, N_{37}=n_{37})}{P(N_{15}+N_{6}=n_{15}+n_{6}, N_{37}=n_{37})}&=\\
        \frac{\frac{N_{tot}!}{n_{6}!n_{15}!n_{37}!(N_{tot}-n_6-n_{15}-n_{37})!}[(1-\phi_{R})(\pi_{\overline{R}1})(\psi_{R})]^{n_6}[(1-\phi_{R})(1-\pi_{\overline{R}1})\psi_{R}]^{n_{15}}[(1-\phi_{R})(1-\psi_R)]^{n_{37}}[\phi_{R}]^{N_{tot}-n_6-n_{15}-n_{37}}}{\frac{N_{tot}!}{(n_{15}+n_{6})!(n_{37}!)(N_{tot}-n_{15}-n_{6}-n_{37})!}[(1-\phi_{R})(\psi_{R})]^{n_{15}+n_{6}}[(1-\phi_{R})(1-\psi_{R})]^{n_{37}}[\phi_{R}]^{N_{tot}-n_{15}-n_{6}-n_{37}}}&=\\
        \frac{(n_{15}+n_{6})!}{(n_{15})!(n_{6})!}[\pi_{\overline{R}1}]^{n_6}[1-\pi_{\overline{R}1}]^{n_{15}}.
    \end{split}
    }
\end{equation*}

 So, $n_6|n_{15}+n_6, n_{15}+n_{6}+n_{37} \sim \text{Binomial}(n_{15}+n_{6}, \pi_{\overline{R}1})$.  Then, $E(\frac{n_6}{n_{15}+n_6}|n_{15}+n_{6}, n_{15}+n_{6}+n_{37})=\pi_{\overline{R}1}$, and the latter term of (A.3) is 0.

The first term in (A.3) is where FPC adjustments are incorporated.  While the multinomial model (Table A1) is convenient for prevalence point estimation, resulting inferences and variance estimates are overly conservative when applied to settings such as CRISP, where the anchor stream is a simple random sample \textit{without replacement}.

To address this issue, and by referring to Table (A1), we can see that if $N_{tot}^{*}=n_{15}+n_{6}+n_{37}$ and $n_{RS}^{*}=n_{15}+n_{6}$ are considered fixed, $\frac{n_{6}}{n_{15}+n_{6}}$ can be viewed as a sample proportion arising from a SRS of size $n_{RS}^{*}$ taken without replacement from a population of size $N_{tot}^{*}$, where $N_{tot}^{*}$ are all the members of the cohort that are not recorded in Stream 1.  From this perspective, we then have that $n_{6}|n_{rs}^{*}, N_{tot}^{*} \sim \text{Hypergeometric}(N_{tot}^{*}, n_{d}^{*}, n_{rs}^{*})$ where $n_{d}^{*}$ are the number of diseased individuals out of $N_{tot}^{*}$.  It then follows from \cite{Cochran} that an unbiased estimate of 
$V(\frac{n_{6}}{n_{15}+n_{6}}|n_{rs}^{*}, N_{tot}^{*})$ is 
\begin{equation}\widehat{V}= \widehat{V}(\frac{n_{6}}{n_{15}+n_{6}}|n_{RS}^{*}, N_{tot}^{*})=\frac{n_{RS}^{*}(N_{tot}^{*}-n_{RS}^{*})}{N_{tot}^{*}(n_{RS}^{*}-1)}\frac{\frac{n_{6}}{n_{15}+n_{6}}(1-\frac{n_{6}}{n_{15}+n_{6}})}{n_{RS}^{*}}.
\end{equation}

That is, $E(\widehat{V}|n_{rs}^{*}, N_{tot}^{*})=V(\frac{n_6}{n_{15}+n_{6}}|n_{rs}^{*}, N_{tot}^{*}).$

Then, 

\begin{equation*}
\begin{split}
    E(\widehat{V})&=E(E[\widehat{V}|n_{RS}^{*}, N_{tot}^{*}])\\
    &=E(V(\frac{n_{6}}{n_{15}+n_{6}}|n_{RS}^{*}, N_{tot}^{*})).
    \end{split}
\end{equation*}

\noindent That is, (A.4) provides unbiased estimation of the first term in (A.3).  As the latter term in (A.3) was shown to be 0, we use (A.4) as the FPC-adjusted variance estimate of $V(\widehat{\pi}_{\overline{R}1})$, referred to as $\widehat{Var}(\widehat{\pi}_{\overline{R}1})_{FPC}$.  The FPC1-adjusted variance of $\widehat{\pi}_5$ is then $\widehat{Var}(\widehat{\pi}_5)_{FPC1}=(1-w)^2\widehat{Var}(\widehat{\pi}_{\overline{R}1})_{FPC}$.

The FPC2-adjusted variance of $\widehat{\pi}_5$ accounts for the variability in $\widehat{\phi}_{R}$. First, we note that under the multinomial model of Table A1, MLEs $\widehat{\phi}_{R}$ and $\widehat{\pi}_{\overline{R}1}$ are asymptotically independent, since $\frac{\partial^2 l}{\partial \phi_{R}\partial\pi_{\overline{R}1}}=0$, where $l$ is the log-likelihood. Then, by the multivariate delta method, we have that,

\begin{equation}
    \begin{split}
\widehat{Var}(\widehat{\pi}_5)&=(1-\widehat{\phi}_R,1-\widehat{\pi}_{\overline{R}1})\begin{pmatrix}  \widehat{Var}(\widehat{\pi}_{\overline{R}1}) & 0 \\
0 & \widehat{Var}(\widehat{\phi}_{R}) \\
\end{pmatrix} \begin{pmatrix}1-\widehat{\phi}_{R} \\ 1-\widehat{\pi}_{\overline{R}1}\end{pmatrix}\\
&=(1-\widehat{\phi}_R)^2\widehat{Var}(\widehat{\pi}_{\overline{R}1})+(1-\widehat{\pi}_{\overline{R}1})^2\widehat{Var}(\widehat{\phi}_{R})\\
    \end{split}
\end{equation}

\noindent where $\widehat{Var}(\widehat{\phi}_{R})=E(-\frac{\partial^2 l}{\partial^2\phi_{R}})^{-1}|_{\phi_{R}=\widehat{\phi}_{R}}=\frac{\widehat{\phi}_R(1-\widehat{\phi}_R)}{N_{tot}}.$

FPC adjustments are then made by plugging $\widehat{Var}(\widehat{\pi}_{\overline{R}1})_{FPC}$ into equation (A.5) for $\widehat{Var}{(\widehat{\pi}_{\overline{R}1}})$, yielding,

\begin{equation}
    \begin{split}
        \widehat{Var}(\widehat{\pi}_5)_{FPC2}&=(1-\widehat{\phi}_R)^2\widehat{Var}(\widehat{\pi}_{\overline{R}1})_{FPC}+(1-\widehat{\pi}_{\overline{R}1})^2\widehat{Var}(\widehat{\phi_{R}})\\
        &=\widehat{Var}(\widehat{\pi}_5)_{FPC1}+\frac{(1-\frac{n_6}{n_{15}+n_6})^2(\frac{n_2+n_4}{N_{tot}})(1-\frac{n_2+n_4}{N_{tot}})}{N_{tot}}.\\
    \end{split}
\end{equation}

However, we find in simulations that FPC2 is generally too conservative and that FPC1 generally performs well.

\subsection{Appendix B: Further Simulation Results}

\renewcommand{\thetable}{B\arabic{table}}
\setcounter{table}{0}


\begin{sidewaystable}[htbp]
\fontsize{8pt}{8pt}\selectfont
\begin{center}
\begin{threeparttable}
\caption{Simulations to Compare Different Estimators of $N$, With $N_{tot}=250$}

\begin{tabular}{|c|c|c|c|c|c|c|c|c|}
\hline
& \multicolumn{2}{|c|}{$\psi=0.1$} & \multicolumn{2}{|c|}{$\psi=0.25$} & \multicolumn{2}{|c|}{$\psi=0.5$} \\
\hline
Estimator & \makecell{Mean (SD) \\ {[avg. SE]}} & \makecell{CI coverage \\ {[avg. width]}} & \makecell{Mean (SD) \\ {[avg. SE]}} & \makecell{CI coverage \\ {[avg. width]}} & \makecell{Mean (SD) \\ {[avg. SE]}} & \makecell{CI coverage \\ {[avg. width]}} \\
\hline
\multicolumn{7}{|c|}{$N=13$}\\
\hline
$\hat{N}_5$ \tnote{[a]} &  \makecell{13.0 (8.6) \\ {[10.2]}, {[9.2]}, {[9.4]}} & \begin{tabular}{ccc}
1.000 & 1.000 & 1.000 \\
{[27.4]} & {[25.4]} & {[25.9]} \\
0.972 & 0.960 &0.961 \\
{[38.0]} & {[34.4]} & {[35.3]} 
\end{tabular} & \makecell{13.1 (4.9) \\ {[6.1]}, {[4.8]}, {[5.3]}} & \begin{tabular}{ccc}
0.952 & 0.895 & 0.927 \\
{[18.3]} & {[15.8]} & {[16.7]} \\
0.986 & 0.918 &0.966 \\
{[23.8]} & {[19.2]} & {[20.9]} 
\end{tabular} & \makecell{13.0 (2.8) \\ {[4.5]}, {[2.8]}, {[3.5]}} & \begin{tabular}{ccc}
0.986 & 0.909 & 0.974 \\
{[13.0]} & {[9.6]} & {[11.0]} \\
0.999 & 0.934 &0.977 \\
{[16.5]} & {[11.2]} & {[13.5]} 
\end{tabular} \\
\hline
$\hat{N}_{RS}$ \tnote{[b]} &  \makecell{13.0 (10.6) \\ {[11.0]}, {[10.6]}} & \begin{tabular}{cc}
0.999 & 0.999 \\
{[29.0]} & {[28.3]} \\
 & 0.971 \\
 & {[39.1]}
\end{tabular} & \makecell{13.1 (6.1) \\ {[6.8]}, {[6.0]}} & \begin{tabular}{cc}
0.883 & 0.884 \\
{[21.2]} & {[20.3]} \\
 & 0.960 \\
 & {[22.2]}
\end{tabular} & \makecell{13.0 (3.5) \\ {[4.9]}, {[3.5]}} & \begin{tabular}{cc}
0.960 & 0.953 \\
{[13.9]} & {[11.3]} \\
 & 0.918 \\
 & {[12.6]}
\end{tabular} \\
\hline
$\hat{N}_{Chap}$ \tnote{[c]}  & \makecell{8.3 (4.6) \\ {[2.9]}} & \makecell{1.000 \\ {[157.3]}} & \makecell{11.2 (5.4) \\ {[4.3]}} & \makecell{0.995 \\ {[93.8]}} & \makecell{12.7 (4.9) \\ {[3.7]}} & \makecell{0.971 \\ {[52.7]}} \\
\hline
\multicolumn{7}{|c|}{$N=25$}\\
\hline
$\hat{N}_5$ \tnote{[a]} &  \makecell{25.1 (11.6) \\ {[12.5]}, {[11.4]}, {[11.8]}} & \begin{tabular}{ccc}
0.892 & 0.846 & 0.877 \\
{[38.6]} & {[36.5]} & {[37.2]} \\
0.979 & 0.972 &0.974 \\
{[47.7]} & {[44.1]} & {[45.4]} 
\end{tabular} & \makecell{25.1 (6.7) \\ {[8.1]}, {[6.6]}, {[7.2]}} & \begin{tabular}{ccc}
0.952 & 0.915 & 0.940 \\
{[27.7]} & {[24.2]} & {[25.5]} \\
0.982 & 0.933 &0.962 \\
{[31.7]} & {[26.1]} & {[28.3]} 
\end{tabular} & \makecell{25.0 (3.9) \\ {[6.1]}, {[3.8]}, {[4.7]}} & \begin{tabular}{ccc}
0.990 & 0.932 & 0.974 \\
{[19.8]} & {[14.7]} & {[17.0]} \\
0.997 & 0.944 &0.982 \\
{[22.9]} & {[15.1]} & {[18.5]} 
\end{tabular} \\
\hline
$\hat{N}_{RS}$ \tnote{[b]} &  \makecell{25.0 (14.3) \\ {[14.4]}, {[13.9]}} & \begin{tabular}{cc}
0.932 & 0.932 \\
{[42.8]} & {[41.8]} \\
 & 0.909 \\
 & {[50.2]}
\end{tabular} & \makecell{25.1 (8.3) \\ {[9.4]}, {[8.2]}} & \begin{tabular}{cc}
0.973 & 0.907 \\
{[30.3]} & {[27.7]} \\
 & 0.954 \\
 & {[30.2]}
\end{tabular} & \makecell{25.0 (4.8) \\ {[6.7]}, {[4.7]}} & \begin{tabular}{cc}
0.982 & 0.937 \\
{[21.0]} & {[16.6]} \\
 & 0.965 \\
 & {[17.5]}
\end{tabular} \\
\hline
$\hat{N}_{Chap}$ \tnote{[c]}  & \makecell{19.4 (9.7) \\ {[7.9]}} & \makecell{1.000 \\ {[141.2]}} & \makecell{24.1 (10.9) \\ {[8.9]}} & \makecell{0.994 \\ {[89.0]}} & \makecell{25.1 (8.0) \\ {[6.1]}} & \makecell{0.971 \\ {[47.7]}} \\
\hline
\multicolumn{7}{|c|}{$N=63$}\\
\hline
$\hat{N}_5$ \tnote{[a]} &  \makecell{63.0 (17.2) \\ {[17.8]}, {[16.8]}, {[17.3]}} & \begin{tabular}{ccc}
0.920 & 0.910 & 0.917 \\
{[65.9]} & {[63.1]} & {[64.4]} \\
0.966 & 0.948 &0.960 \\
{[67.7]} & {[64.0]} & {[65.8]} 
\end{tabular} & \makecell{63.0 (9.8) \\ {[11.9]}, {[9.8]}, {[10.6]}} & \begin{tabular}{ccc}
0.973 & 0.938 & 0.955 \\
{[46.4]} & {[38.6]} & {[41.4]} \\
0.981 & 0.949 &0.965 \\
{[46.1]} & {[38.2]} & {[41.1]} 
\end{tabular} & \makecell{63.1 (5.6) \\ {[8.9]}, {[5.7]}, {[6.9]}} & \begin{tabular}{ccc}
0.997 & 0.948 & 0.980 \\
{[34.5]} & {[22.3]} & {[26.9]} \\
0.998 & 0.951 &0.982 \\
{[34.6]} & {[22.1]} & {[26.8]} 
\end{tabular} \\
\hline
$\hat{N}_{RS}$ \tnote{[b]} & \makecell{63.0 (20.8) \\ {[21.1]}, {[20.5]}} & \begin{tabular}{cc}
0.908 & 0.908 \\
{[73.0]} & {[71.3]} \\
 & 0.949 \\
 & {[73.4]}
\end{tabular} & \makecell{62.9 (11.9) \\ {[13.7]}, {[11.9]}} & \begin{tabular}{cc}
0.978 & 0.953 \\
{[51.1]} & {[45.4]} \\
 & 0.958 \\
 & {[45.4]}
\end{tabular} & \makecell{63.1 (6.8) \\ {[9.7]}, {[6.9]}} & \begin{tabular}{cc}
0.989 & 0.945 \\
{[36.3]} & {[26.7]} \\
 & 0.961 \\
 & {[26.6]}
\end{tabular} \\
\hline
$\hat{N}_{Chap}$ \tnote{[c]}  & \makecell{60.3 (29.1) \\ {[23.7]}} & \makecell{0.993 \\ {[143.8]}} & \makecell{62.9 (20.0) \\ {[16.5]}} & \makecell{0.969 \\ {[90.8]}} & \makecell{63.1 (10.5) \\ {[9.8]}} & \makecell{0.960 \\ {[50.6]}} \\
\hline
\multicolumn{7}{|c|}{$N=125$}\\
\hline
$\hat{N}_5$ \tnote{[a]} & \makecell{125.2 (20.7) \\ {[21.6]}, {[20.7]}, {[21.1]}} & \begin{tabular}{ccc}
0.948 & 0.936 & 0.941 \\
{[84.6]} & {[81.1]} & {[82.5]} \\
0.963 & 0.952 &0.957 \\
{[80.4]} & {[76.9]} & {[78.3]} 
\end{tabular} & \makecell{124.8 (11.9) \\ {[14.3]}, {[12.0]}, {[12.6]}} & \begin{tabular}{ccc}
0.977 & 0.943 & 0.958 \\
{[55.9]} & {[47.0]} & {[49.5]} \\
0.982 & 0.950 &0.964 \\
{[54.6]} & {[46.0]} & {[48.4]} 
\end{tabular} & \makecell{125.0 (6.9) \\ {[10.5]}, {[6.9]}, {[7.9]}} & \begin{tabular}{ccc}
0.997 & 0.947 & 0.973 \\
{[41.0]} & {[27.0]} & {[31.1]} \\
0.997 & 0.947 &0.974 \\
{[40.5]} & {[26.7]} & {[30.7]} 
\end{tabular} \\
\hline
$\hat{N}_{RS}$ \tnote{[b]} &  \makecell{125.4 (23.7) \\ {[24.5]}, {[23.8]}} & \begin{tabular}{cc}
0.968 & 0.913 \\
{[94.1]} & {[91.4]} \\
 & 0.968 \\
 & {[88.0]}
\end{tabular} & \makecell{124.9 (13.6) \\ {[15.8]}, {[13.8]}} & \begin{tabular}{cc}
0.976 & 0.947 \\
{[61.8]} & {[54.1]} \\
 & 0.947 \\
 & {[53.0]}
\end{tabular} & \makecell{125.0 (8.0) \\ {[11.2]}, {[7.9]}} & \begin{tabular}{cc}
0.993 & 0.958 \\
{[43.7]} & {[31.1]} \\
 & 0.958 \\
 & {[30.8]}
\end{tabular} \\
\hline
$\hat{N}_{Chap}$ \tnote{[c]} & \makecell{124.7 (49.8) \\ {[38.7]}} & \makecell{0.975 \\ {[137.3]}} & \makecell{124.7 (25.2) \\ {[23.4]}} & \makecell{0.961 \\ {[95.9]}} & \makecell{124.9 (14.2) \\ {[13.7]}} & \makecell{0.958 \\ {[58.0]}} \\
\hline
\end{tabular}

\begin{tablenotes}

\item[a] Reported standard errors are the unadjusted (left), the FPC1-adjusted (center), and the FPC2-adjusted (right); Confidence intervals are a Wald unadjusted (top left), Wald FPC1-adjusted (top center) Wald-FPC2 adjusted (top right), unadjusted Bayesian credible interval (bottom left), FPC1-adjusted Bayesian credible interval (bottom center), and FPC2-adjusted Bayesian credible interval (bottom right).
\item[b] Reported standard errors are the non-FPC-adjusted (left) and Cochran's FPC-adjusted (right); Confidence intervals are non-FPC-adjusted Wald (top left), Cochran's FPC-adjusted Wald (top right), and a FPC-adjusted Jeffreys prior-based credible interval (bottom right) (\cite{Lyles2023}).

\item[c] Reported confidence interval is a logit-based confidence interval given by \cite{Sadinle2009}.
\end{tablenotes}
\end{threeparttable}
\end{center}
\end{sidewaystable}

\clearpage
\newpage

\begin{sidewaystable}[p]
\fontsize{8pt}{8pt}\selectfont
\begin{center}
\begin{threeparttable}
\caption{Simulations to Compare Different Estimators of $N$, With $N_{tot}=500$}


\begin{tabular}{|c|c|c|c|c|c|c|c|c|}
\hline
& \multicolumn{2}{|c|}{$\psi=0.1$} & \multicolumn{2}{|c|}{$\psi=0.25$} & \multicolumn{2}{|c|}{$\psi=0.5$} \\
\hline
Estimator & \makecell{Mean (SD) \\ {[avg. SE]}} & \makecell{CI coverage \\ {[avg. width]}} & \makecell{Mean (SD) \\ {[avg. SE]}} & \makecell{CI coverage \\ {[avg. width]}} & \makecell{Mean (SD) \\ {[avg. SE]}} & \makecell{CI coverage \\ {[avg. width]}} \\
\hline
\multicolumn{7}{|c|}{$N=25$}\\
\hline
$\hat{N}_5$ \tnote{[a]}  & \makecell{25.0 (11.7) \\ {[12.9]}, {[11.5]}, {[11.9]}} & \begin{tabular}{ccc}
0.888 & 0.843 & 0.880 \\
{[39.3]} & {[36.7]} & {[37.5]} \\
0.979 & 0.968 &0.971 \\
{[50.4]} & {[45.5]} & {[47.0]} 
\end{tabular} & \makecell{25.0 (6.8) \\ {[8.2]}, {[6.6]}, {[7.3]}} & \begin{tabular}{ccc}
0.951 & 0.908 & 0.934 \\
{[27.9]} & {[24.3]} & {[25.7]} \\
0.981 & 0.928 &0.959 \\
{[32.7]} & {[26.6]} & {[29.1]} 
\end{tabular} & \makecell{25.0 (3.9) \\ {[6.2]}, {[3.9]}, {[4.9]}} & \begin{tabular}{ccc}
0.991 & 0.934 & 0.976 \\
{[20.1]} & {[14.9]} & {[17.3]} \\
0.999 & 0.943 &0.981 \\
{[23.5]} & {[15.5]} & {[19.1]} 
\end{tabular} \\
\hline
$\hat{N}_{RS}$ \tnote{[b]} &  \makecell{24.8 (14.6) \\ {[14.7]}, {[14.1]}} & \begin{tabular}{cc}
0.930 & 0.930 \\
{[43.8]} & {[42.5]} \\
 & 0.902 \\
 & {[52.9]}
\end{tabular} & \makecell{25.0 (8.5) \\ {[9.6]}, {[8.3]}} & \begin{tabular}{cc}
0.970 & 0.906 \\
{[31.3]} & {[28.8]} \\
 & 0.945 \\
 & {[31.2]}
\end{tabular} & \makecell{25.0 (4.9) \\ {[6.9]}, {[4.9]}} & \begin{tabular}{cc}
0.982 & 0.944 \\
{[21.3]} & {[16.9]} \\
 & 0.964 \\
 & {[18.1]}
\end{tabular} \\
\hline
$\hat{N}_{Chap}$ \tnote{[c]} & \makecell{19.3 (9.7) \\ {[7.9]}} & \makecell{1.000 \\ {[228.8]}} & \makecell{24.0 (10.9) \\ {[8.8]}} & \makecell{0.995 \\ {[109.3]}} & \makecell{25.0 (7.5) \\ {[6.0]}} & \makecell{0.973 \\ {[52.5]}} \\
\hline
\multicolumn{7}{|c|}{$N=50$}\\
\hline
$\hat{N}_5$ \tnote{[a]} &  \makecell{49.9 (16.3) \\ {[17.1]}, {[15.8]}, {[16.3]}} & \begin{tabular}{ccc}
0.911 & 0.870 & 0.892 \\
{[60.6]} & {[57.5]} & {[58.7]} \\
0.949 & 0.937 &0.943 \\
{[67.9]} & {[62.8]} & {[64.9]} 
\end{tabular} & \makecell{50.0 (9.3) \\ {[11.4]}, {[9.3]}, {[10.1]}} & \begin{tabular}{ccc}
0.970 & 0.934 & 0.954 \\
{[43.6]} & {[36.4]} & {[39.2]} \\
0.983 & 0.946 &0.965 \\
{[45.2]} & {[36.9]} & {[40.1]} 
\end{tabular} & \makecell{50.1 (5.4) \\ {[8.6]}, {[5.4]}, {[6.7]}} & \begin{tabular}{ccc}
0.997 & 0.946 & 0.983 \\
{[32.3]} & {[21.3]} & {[26.1]} \\
0.999 & 0.949 &0.983 \\
{[33.8]} & {[21.4]} & {[26.4]} 
\end{tabular} \\
\hline
$\hat{N}_{RS}$ \tnote{[b]} &  \makecell{50.1 (20.3) \\ {[20.7]}, {[19.8]}} & \begin{tabular}{cc}
0.897 & 0.897 \\
{[69.4]} & {[67.3]} \\
 & 0.953 \\
 & {[73.5]}
\end{tabular} & \makecell{50.0 (11.6) \\ {[13.3]}, {[11.6]}} & \begin{tabular}{cc}
0.957 & 0.951 \\
{[47.7]} & {[42.7]} \\
 & 0.964 \\
 & {[44.2]}
\end{tabular} & \makecell{50.0 (6.7) \\ {[9.5]}, {[6.7]}} & \begin{tabular}{cc}
0.988 & 0.940 \\
{[33.7]} & {[25.5]} \\
 & 0.951 \\
 & {[25.9]}
\end{tabular} \\
\hline
$\hat{N}_{Chap}$ \tnote{[c]} & \makecell{45.4 (22.1) \\ {[18.6]}} & \makecell{0.996 \\ {[206.4]}} & \makecell{49.7 (18.2) \\ {[14.7]}} & \makecell{0.977 \\ {[106.2]}} & \makecell{50.1 (10.1) \\ {[8.9]}} & \makecell{0.964 \\ {[52.4]}} \\
\hline
\multicolumn{7}{|c|}{$N=125$}\\
\hline
$\hat{N}_5$ \tnote{[a]} & \makecell{125.0 (24.1) \\ {[25.5]}, {[23.9]}, {[24.5]}} & \begin{tabular}{ccc}
0.947 & 0.930 & 0.939 \\
{[99.4]} & {[93.4]} & {[95.7]} \\
0.962 & 0.947 &0.954 \\
{[98.6]} & {[92.3]} & {[94.8]} 
\end{tabular} & \makecell{124.9 (13.8) \\ {[16.8]}, {[13.8]}, {[14.9]}} & \begin{tabular}{ccc}
0.979 & 0.941 & 0.961 \\
{[66.0]} & {[54.3]} & {[58.4]} \\
0.983 & 0.951 &0.968 \\
{[65.6]} & {[53.9]} & {[58.0]} 
\end{tabular} & \makecell{124.8 (8.1) \\ {[12.5]}, {[8.0]}, {[9.7]}} & \begin{tabular}{ccc}
0.997 & 0.941 & 0.980 \\
{[49.2]} & {[31.4]} & {[38.0]} \\
0.998 & 0.945 &0.980 \\
{[49.0]} & {[31.3]} & {[37.8]} 
\end{tabular} \\
\hline
$\hat{N}_{RS}$ \tnote{[b]} &  \makecell{125.2 (29.1) \\ {[30.3]}, {[29.0]}} & \begin{tabular}{cc}
0.954 & 0.941 \\
{[113.9]} & {[109.7]} \\
 & 0.963 \\
 & {[109.7]}
\end{tabular} & \makecell{124.8 (16.7) \\ {[19.3]}, {[16.8]}} & \begin{tabular}{cc}
0.976 & 0.956 \\
{[75.3]} & {[65.6]} \\
 & 0.946 \\
 & {[65.1]}
\end{tabular} & \makecell{124.8 (9.8) \\ {[13.7]}, {[9.7]}} & \begin{tabular}{cc}
0.992 & 0.945 \\
{[53.4]} & {[38.0]} \\
 & 0.945 \\
 & {[37.8]}
\end{tabular} \\
\hline
$\hat{N}_{Chap}$ \tnote{[c]} & \makecell{124.3 (52.0) \\ {[39.6]}} & \makecell{0.977 \\ {[217.5]}} & \makecell{125.1 (26.2) \\ {[24.2]}} & \makecell{0.956 \\ {[119.9]}} & \makecell{124.8 (14.5) \\ {[14.0]}} & \makecell{0.959 \\ {[64.2]}} \\
\hline
\multicolumn{7}{|c|}{$N=250$}\\
\hline
$\hat{N}_5$ \tnote{[a]} &  \makecell{249.9 (29.4) \\ {[31.0]}, {[29.3]}, {[29.8]}} & \begin{tabular}{ccc}
0.952 & 0.939 & 0.943 \\
{[121.5]} & {[114.9]} & {[116.9]} \\
0.963 & 0.949 &0.955 \\
{[118.1]} & {[111.6]} & {[113.7]} 
\end{tabular} & \makecell{249.7 (16.8) \\ {[20.2]}, {[16.9]}, {[17.8]}} & \begin{tabular}{ccc}
0.979 & 0.951 & 0.961 \\
{[79.1]} & {[66.3]} & {[69.7]} \\
0.981 & 0.952 &0.962 \\
{[78.2]} & {[65.5]} & {[68.9]} 
\end{tabular} & \makecell{250.0 (9.8) \\ {[14.8]}, {[9.8]}, {[11.2]}} & \begin{tabular}{ccc}
0.996 & 0.952 & 0.973 \\
{[58.1]} & {[38.3]} & {[43.9]} \\
0.997 & 0.951 &0.974 \\
{[57.7]} & {[38.0]} & {[43.6]} 
\end{tabular} \\
\hline
$\hat{N}_{RS}$ \tnote{[b]} & \makecell{250.0 (33.5) \\ {[35.0]}, {[33.6]}} & \begin{tabular}{cc}
0.948 & 0.948 \\
{[137.2]} & {[131.5]} \\
 & 0.948 \\
 & {[128.5]}
\end{tabular} & \makecell{249.8 (19.4) \\ {[22.3]}, {[19.4]}} & \begin{tabular}{cc}
0.978 & 0.936 \\
{[87.4]} & {[76.0]} \\
 & 0.961 \\
 & {[75.3]}
\end{tabular} & \makecell{249.9 (11.1) \\ {[15.8]}, {[11.2]}} & \begin{tabular}{cc}
0.994 & 0.942 \\
{[61.9]} & {[43.9]} \\
 & 0.942 \\
 & {[43.7]}
\end{tabular} \\
\hline
$\hat{N}_{Chap}$ \tnote{[c]} & \makecell{249.8 (65.2) \\ {[57.7]}} & \makecell{0.959 \\ {[230.8]}} & \makecell{249.5 (35.2) \\ {[33.9]}} & \makecell{0.958 \\ {[144.7]}} & \makecell{250.2 (20.1) \\ {[19.8]}} & \makecell{0.956 \\ {[83.5]}} \\
\hline
\end{tabular}

\begin{tablenotes}

\item[a] Reported standard errors are the unadjusted (left), the FPC1-adjusted (center), and the FPC2-adjusted (right); Confidence intervals are a Wald unadjusted (top left), Wald FPC1-adjusted (top center) Wald-FPC2 adjusted (top right), unadjusted Bayesian credible interval (bottom left), FPC1-adjusted Bayesian credible interval (bottom center), and FPC2-adjusted Bayesian credible interval (bottom right).
\item[b] Reported standard errors are the non-FPC-adjusted (left) and Cochran's FPC-adjusted (right); Confidence intervals are non-FPC-adjusted Wald (top left), Cochran's FPC-adjusted Wald (top right), and a FPC-adjusted Jeffreys prior-based credible interval (bottom right) (\cite{Lyles2023}).

\item[c] Reported confidence interval is a logit-based confidence interval given by \cite{Sadinle2009}.
\end{tablenotes}
\end{threeparttable}
\end{center}
\end{sidewaystable}

\clearpage
\newpage


\newpage
\clearpage
\begin{sidewaystable}[p]
\fontsize{8pt}{8pt}\selectfont
\begin{center}
\begin{threeparttable}
\caption{Simulations to Compare Different Estimators of $N$ in a Population Similar to the  CRISP Cohort\tnote{*}}
\begin{tabular}{|c|c|c|c|c|c|c|}
\hline
& \multicolumn{2}{|c|}{$p_{1,\,symp}=0.50$} & \multicolumn{2}{|c|}{$p_{1,\,symp}=0.75$} & \multicolumn{2}{|c|}{$p_{1,\,symp}=0.90$} \\
\hline
Estimator & \makecell{Mean (SD) \\ {[avg. SE]}} & \makecell{CI coverage \\ {[avg. width]}} & \makecell{Mean (SD) \\ {[avg. SE]}} & \makecell{CI coverage \\ {[avg. width]}} & \makecell{Mean (SD) \\ {[avg. SE]}} & \makecell{CI coverage \\ {[avg. width]}} \\
\hline
\multicolumn{7}{|c|}{$p_{symp,flu}=0.25$}\\
\hline
$\hat{N}_5$ \tnote{[a]} & \makecell{155.8 (20.3) \\ {[23.3]}, {[20.3]}, {[21.1]}} & \begin{tabular}{ccc}
0.968 & 0.942 & 0.951 \\
{[91.2]} & {[79.6]} & {[82.7]} \\
0.974 & 0.950 &0.957 \\
{[91.1]} & {[79.5]} & {[82.6]} 
\end{tabular} & \makecell{155.7 (19.9) \\ {[22.6]}, {[19.5]}, {[20.5]}} & \begin{tabular}{ccc}
0.968 & 0.939 & 0.953 \\
{[88.4]} & {[76.3]} & {[80.3]} \\
0.974 & 0.945 &0.957 \\
{[88.4]} & {[76.3]} & {[80.3]} 
\end{tabular} & \makecell{155.8 (18.9) \\ {[22.1]}, {[19.0]}, {[20.1]}} & \begin{tabular}{ccc}
0.968 & 0.942 & 0.955 \\
{[86.8]} & {[74.4]} & {[78.9]} \\
0.979 & 0.947 &0.963 \\
{[86.8]} & {[74.4]} & {[78.9]} 
\end{tabular} \\
\hline
$\hat{N}_{RS}$ \tnote{[b]} & \makecell{155.9 (23.4) \\ {[26.0]}, {[23.4]}} & \begin{tabular}{cc}
0.969 & 0.947 \\
{[101.8]} & {[91.6]} \\
 & 0.954 \\
 & {[91.4]}
\end{tabular} & \makecell{155.8 (23.6) \\ {[26.0]}, {[23.4]}} & \begin{tabular}{cc}
0.967 & 0.945 \\
{[101.6]} & {[91.5]} \\
 & 0.954 \\
 & {[91.3]}
\end{tabular} & \makecell{155.7 (23.3) \\ {[26.0]}, {[23.4]}} & \begin{tabular}{cc}
0.969 & 0.947 \\
{[101.4]} & {[91.4]} \\
 & 0.954 \\
 & {[91.3]}
\end{tabular} \\
\hline
$\hat{N}_{Chap}$  \tnote{[c]}& \makecell{155.5 (46.3) \\ {[39.8]}} & \makecell{0.963 \\ {[213.5]}} & \makecell{155.1 (37.4) \\ {[34.1]}} & \makecell{0.960 \\ {[175.6]}} & \makecell{156.0 (35.4) \\ {[32.0]}} & \makecell{0.960 \\ {[161.3]}} \\
\hline
\multicolumn{7}{|c|}{$p_{symp,flu}=0.50$}\\
\hline
$\hat{N}_5$ \tnote{[a]} & \makecell{156.0 (19.3) \\ {[22.5]}, {[19.4]}, {[20.5]}} & \begin{tabular}{ccc}
0.972 & 0.943 & 0.955 \\
{[88.3]} & {[76.2]} & {[80.2]} \\
0.977 & 0.950 &0.961 \\
{[88.3]} & {[76.2]} & {[80.2]} 
\end{tabular} & \makecell{155.9 (18.0) \\ {[21.1]}, {[17.7]}, {[19.3]}} & \begin{tabular}{ccc}
0.972 & 0.939 & 0.958 \\
{[82.7]} & {[69.5]} & {[75.5]} \\
0.979 & 0.944 &0.964 \\
{[82.8]} & {[69.6]} & {[75.6]} 
\end{tabular} & \makecell{155.7 (16.7) \\ {[20.1]}, {[16.6]}, {[18.5]}} & \begin{tabular}{ccc}
0.973 & 0.936 & 0.961 \\
{[78.9]} & {[65.0]} & {[72.4]} \\
0.984 & 0.946 &0.973 \\
{[79.3]} & {[65.3]} & {[72.7]} 
\end{tabular} \\
\hline
$\hat{N}_{RS}$ \tnote{[b]} & \makecell{156.1 (23.2) \\ {[26.0]}, {[23.4]}} & \begin{tabular}{cc}
0.972 & 0.948 \\
{[101.7]} & {[91.6]} \\
 & 0.955 \\
 & {[91.4]}
\end{tabular} & \makecell{156.1 (23.3) \\ {[26.0]}, {[23.4]}} & \begin{tabular}{cc}
0.970 & 0.947 \\
{[100.2]} & {[90.7]} \\
 & 0.954 \\
 & {[90.9]}
\end{tabular} & \makecell{155.9 (23.5) \\ {[26.0]}, {[23.4]}} & \begin{tabular}{cc}
0.967 & 0.944 \\
{[97.7]} & {[88.9]} \\
 & 0.951 \\
 & {[89.7]}
\end{tabular} \\
\hline
$\hat{N}_{Chap}$ \tnote{[c]} & \makecell{155.3 (36.2) \\ {[33.9]}} & \makecell{0.962 \\ {[174.3]}} & \makecell{155.7 (28.5) \\ {[26.7]}} & \makecell{0.959 \\ {[130.5]}} & \makecell{155.6 (24.5) \\ {[23.1]}} & \makecell{0.956 \\ {[111.8]}} \\
\hline
\multicolumn{7}{|c|}{$p_{symp,flu}=0.75$}\\
\hline
$\hat{N}_5$ \tnote{[a]} & \makecell{156.1 (18.5) \\ {[21.6]}, {[18.3]}, {[19.7]}} & \begin{tabular}{ccc}
0.972 & 0.940 & 0.958 \\
{[84.5]} & {[71.7]} & {[77.0]} \\
0.979 & 0.945 &0.961 \\
{[84.6]} & {[71.7]} & {[77.1]} 
\end{tabular} & \makecell{156.2 (15.4) \\ {[19.1]}, {[15.2]}, {[17.6]}} & \begin{tabular}{ccc}
0.979 & 0.939 & 0.969 \\
{[74.3]} & {[59.6]} & {[68.7]} \\
0.985 & 0.945 &0.976 \\
{[75.1]} & {[60.0]} & {[69.3]} 
\end{tabular} & \makecell{156.1 (12.9) \\ {[17.3]}, {[12.9]}, {[16.1]}} & \begin{tabular}{ccc}
0.986 & 0.929 & 0.981 \\
{[65.0]} & {[50.3]} & {[61.2]} \\
0.990 & 0.944 &0.985 \\
{[67.7]} & {[51.0]} & {[63.3]} 
\end{tabular} \\
\hline
$\hat{N}_{RS}$ \tnote{[b]} & \makecell{156.1 (23.6) \\ {[26.0]}, {[23.4]}} & \begin{tabular}{cc}
0.968 & 0.944 \\
{[100.9]} & {[91.1]} \\
 & 0.952 \\
 & {[91.2]}
\end{tabular} & \makecell{156.3 (23.4) \\ {[26.0]}, {[23.4]}} & \begin{tabular}{cc}
0.969 & 0.946 \\
{[93.3]} & {[85.5]} \\
 & 0.952 \\
 & {[87.2]}
\end{tabular} & \makecell{156.1 (23.2) \\ {[26.0]}, {[23.4]}} & \begin{tabular}{cc}
0.971 & 0.949 \\
{[83.4]} & {[77.0]} \\
 & 0.956 \\
 & {[80.1]}
\end{tabular} \\
\hline
$\hat{N}_{Chap}$ \tnote{[c]} & \makecell{156.1 (32.0) \\ {[28.9]}} & \makecell{0.957 \\ {[143.2]}} & \makecell{156.1 (20.8) \\ {[19.7]}} & \makecell{0.959 \\ {[94.9]}} & \makecell{156.1 (16.1) \\ {[15.3]}} & \makecell{0.958 \\ {[74.4]}} \\
\hline
\multicolumn{7}{|c|}{$p_{symp,flu}=0.90$}\\
\hline
$\hat{N}_5$ \tnote{[a]} & \makecell{155.9 (17.6) \\ {[20.9]}, {[17.5]}, {[19.1]}} & \begin{tabular}{ccc}
0.975 & 0.940 & 0.962 \\
{[82.0]} & {[68.7]} & {[74.9]} \\
0.982 & 0.948 &0.968 \\
{[82.2]} & {[68.8]} & {[75.1]} 
\end{tabular} & \makecell{156.2 (13.3) \\ {[17.6]}, {[13.4]}, {[16.4]}} & \begin{tabular}{ccc}
0.985 & 0.934 & 0.980 \\
{[67.1]} & {[52.2]} & {[62.9]} \\
0.990 & 0.944 &0.983 \\
{[69.3]} & {[52.9]} & {[64.6]} 
\end{tabular} & \makecell{156.1 (10.1) \\ {[15.1]}, {[9.8]}, {[14.4]}} & \begin{tabular}{ccc}
0.994 & 0.910 & 0.993 \\
{[49.5]} & {[36.9]} & {[47.8]} \\
0.997 & 0.924 &0.994 \\
{[55.0]} & {[38.6]} & {[52.7]} 
\end{tabular} \\
\hline
$\hat{N}_{RS}$ \tnote{[b]} & \makecell{156.0 (23.3) \\ {[26.0]}, {[23.4]}} & \begin{tabular}{cc}
0.971 & 0.949 \\
{[99.8]} & {[90.4]} \\
 & 0.955 \\
 & {[90.7]}
\end{tabular} & \makecell{155.9 (23.4) \\ {[26.0]}, {[23.4]}} & \begin{tabular}{cc}
0.970 & 0.950 \\
{[85.4]} & {[78.9]} \\
 & 0.954 \\
 & {[81.5]}
\end{tabular} & \makecell{156.5 (23.4) \\ {[26.0]}, {[23.4]}} & \begin{tabular}{cc}
0.972 & 0.952 \\
{[72.1]} & {[67.5]} \\
 & 0.954 \\
 & {[70.3]}
\end{tabular} \\
\hline
$\hat{N}_{Chap}$ \tnote{[c]} & \makecell{155.7 (27.9) \\ {[26.0]}} & \makecell{0.961 \\ {[126.9]}} & \makecell{156.4 (17.1) \\ {[16.2]}} & \makecell{0.954 \\ {[78.6]}} & \makecell{156.0 (11.3) \\ {[10.6]}} & \makecell{0.952 \\ {[55.8]}} \\
\hline
\end{tabular} 
\begin{tablenotes}
\item[*] To mimic the CRISP cohort, we set $N_{tot}=1029$, $N=156$, and the size of the anchor stream to $n_{RS}=200$.

\item[a] Reported standard errors are the unadjusted (left), the FPC1-adjusted (center), and the FPC2-adjusted (right); Confidence intervals are a Wald unadjusted (top left), Wald FPC1-adjusted (top center) Wald-FPC2 adjusted (top right), unadjusted Bayesian credible interval (bottom left), FPC1-adjusted Bayesian credible interval (bottom center), and FPC2-adjusted Bayesian credible interval (bottom right).
\item[b] Reported standard errors are the non-FPC-adjusted (left) and Cochran's FPC-adjusted (right); Confidence intervals are non-FPC-adjusted Wald (top left), Cochran's FPC-adjusted Wald (top right), and a FPC-adjusted Jeffreys prior-based credible interval (bottom right) (\cite{Lyles2023}).

\item[c] Reported confidence interval is a logit-based confidence interval given by \cite{Sadinle2009}.
\end{tablenotes}
\end{threeparttable}
\end{center}
\end{sidewaystable}

\clearpage
\newpage

\end{document}